\documentclass[11pt,epsfig,letterpaper]{article}
\usepackage{varioref,exscale,latexsym,amsmath,amssymb}
\usepackage{graphicx}
\def\beq{\begin{equation}}

\def\eeq{\end{equation}}

\def\beqa{\begin{eqnarray}}

\def\eeqa{\end{eqnarray}}

\title{{\bf Quark and lepton masses and mixing in the landscape}}
\author{John F. Donoghue, Koushik Dutta and Andreas Ross \\ \\
 Department of Physics\\
University of Massachusetts\\
Amherst, MA  01003, USA\\  \\}

\begin{document}
\maketitle
\thispagestyle{empty}
\begin{abstract}
Even if quark and lepton masses are not uniquely predicted by the
fundamental theory, as may be the case in the string theory
landscape, nevertheless their pattern may reveal features of the
underlying theory. We use statistical techniques to show that the
observed masses appear to be representative of a scale invariant
distribution, $\rho(m)\sim1/m$. If we extend this distribution to
include all the Yukawa couplings, we show that the resulting CKM
matrix elements typically show a hierarchical pattern similar to
observations. The Jarlskog invariant measuring the amount of CP
violation is also well reproduced in magnitude. We also apply this
framework to neutrinos using the seesaw mechanism. The neutrino
results are ambiguous, with the observed pattern being statistically
allowed even though the framework does not provide a natural
explanation for the observed two large mixing angles.
Our framework highly favors a normal hierarchy of neutrino masses.
We also are able to make statistical predictions in the neutrino
sector when we specialize to situations consistent with the known
mass differences and two large mixing angles. Within our framework,
we show that with 95\% confidence the presently unmeasured MNS mixing angle
$\sin \theta_{13}$ is larger than 0.04 and typically of order 0.1.
The leptonic Jarlskog invariant is found to be typically of order
$10^{-2}$ and the magnitude of the effective Majorana mass $m_{ee}$
is typically of order 0.001 eV.

\end{abstract}
\vspace{0.2 in}
\setcounter{page}{0}
\newpage

\section{Introduction}

Of the 28 parameters of the Standard Model (including neutrino
masses and mixing) the quark and lepton masses appear particularly
puzzling. We would expect that the spectrum of masses would exhibit
some underlying structure, much as the periodic table and the hadron
spectrum respectively reveal the dynamics of atoms and elementary
particles. However, apparently the quark and lepton masses do not
show any such pattern. Aside from a rough correlation of mass with
generation, decades of searching have not revealed any significant
regularity in the masses. Perhaps there is none.

In this paper, we explore the possibility that the masses and
mixings are not uniquely predicted, but are representative of an
ensemble of possibilities. The masses and mixing then reveal the
``weight'' or ``measure'' of the underlying theory. This possibility
was first proposed by Donoghue in \cite{weight}, and we develop it
further. This is a plausible outcome of the string theory landscape
\cite{landscape}. In this description, there are very many Standard
Model vacua with parameters close to those observed, and yet many
more with different parameters. While we observe only a single
ground state, the solution that we observe is representative of the
ensemble of possible SM vacua that occur in the landscape. In
particular, the Yukawa couplings would not be unique, but would be
representative of the couplings found in the ensemble of solutions.
Since we have many manifestations of the Yukawa couplings (9 quark
and lepton masses, two neutrino mass differences as well as the CKM
and MNS weak mixing elements) we may apply statistical methods even
though we live in a single member of the possible landscape vacua.
Our primary assumption is that the observed masses are
representative of this ensemble. By studying the phenomenology of
the masses, we then learn about the nature of the underlying theory.

\begin{figure}[ht]
 \begin{center}
  \includegraphics[scale=0.75]{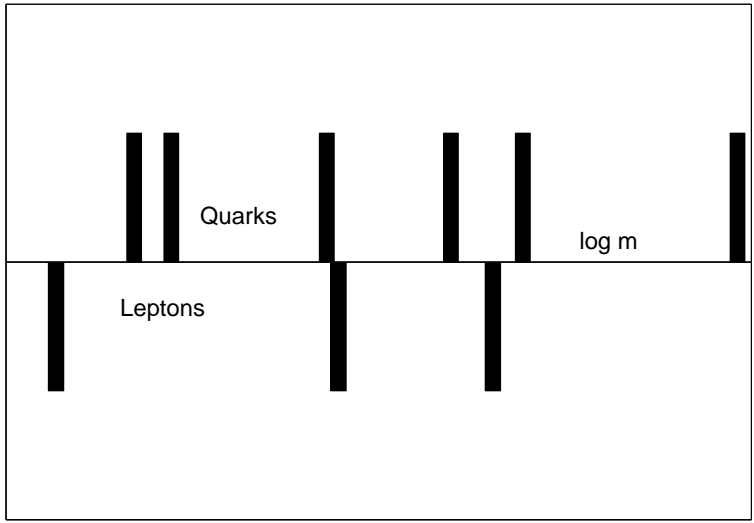}
 \end{center}
 \caption{\small{Quark and lepton masses, defined at the energy $\mu=M_W$, on a logarithmic scale. A
 scale invariant weight corresponds to a uniform distribution on this scale.}}
 \label{scaleinvariant}
\end{figure}

By employing statistical tests we will show that the observed quark
masses appear to be distributed in a scale-invariant fashion. This was
already suggested in \cite{weight} but we provide better statistical measures.
This result can be readily seen in Fig. \ref{scaleinvariant}. A scale
invariant set of masses is one where the values appear as a uniform random
distribution when plotted on a log scale. The masses of the quarks and
charged leptons are shown on such a scale in Fig. \ref{scaleinvariant} and
appear visually to be consistent with being uniformly random. In Section 3,
we use statistical tests to confirm this and to quantify how far the weight
could deviate from a scale invariant form. In Section 4 we apply this
distribution to the full Yukawa matrices of quarks and investigate the
weak mixing angles. We will see that one naturally generates a hierarchy
in the CKM elements similar to the one found in nature, aside from a
discrete ambiguity concerning the generation structure.
Moreover, the magnitude of the observed CP violation is readily reproduced.

The neutrino masses appear quite different from other lepton masses,
with a possible explanation being the seesaw mechanism and the
presence of heavy right handed Majorana mass. We explore the
neutrino masses in a similar fashion, using the same weight for the
Dirac masses and allowing a different weight for the Majorana
masses. We describe the neutrino phenomenology in Section 5.

It is possible that these results could be useful in testing or
exploring the string landscape. For example, it has been
suggested \cite{Mdynamics} that a scale invariant weight could emerge
from the intersecting brane worlds construction \cite{ibw} of the
Standard Model. If some branches of the landscape yield weights
compatible with observations while others do not, this could be used
to refine our location in the space of string theory solutions. We
are clearly far from an understanding of the landscape, but the
masses and mixings of the leptons can play a useful role in
landscape phenomenology.

\section{The properties of the weight}

The most fundamental assumption of our approach is the existence of
a large ensemble of nearly equivalent vacuum states with different
values of the quark masses, of which our vacua is assumed to be
representative. The string landscape could be such a theory. It has
been estimated \cite{landscape} that there are
perhaps $10^{100}$ Standard Model vacua with parameters agreeing
with observation within the experimental error bars. There are then
many more that look like the Standard Model but have other values of
the quark and lepton masses. This larger set of SM vacua then
constitutes our ensemble.

Within this ensemble we need to make other assumptions. For example,
we are assuming that the quark masses are not correlated with each
other. If there are really of order $10^{100}$ vacua this seems
reasonable. We could hold any one of the quark masses fixed and
still find a whole range of values for the other masses. We also
will make the assumption that quark and charged lepton masses (but
not neutrino masses) play by the same rules, although we will
explore some slight differences that arise due to the differences
between quarks and leptons. Finally, we need to make assumptions
about the whole matrix of Yukawa couplings, as will be discussed
below.

The weight $\rho(m) $ is a probability distribution function for the
masses (or later for the Yukawa couplings). It is defined by
considering the fraction of masses $f$ found at a value $m$ within
$dm$ as
\begin{equation}
f(m) = \rho(m) dm.
\end{equation}
The normalization of the weight is by definition then
\begin{equation}
1=\int \rho (m) dm.
\end{equation}
For the simple weights $\rho (m)$ that we consider, this
normalization condition demands upper limits and,in some cases,
lower limits for the range of allowed masses.

Because the masses depend on the energy scale, the weight will also
be depend on this scale. The renormalization group allows us to
determine the scale dependence, as was shown in \cite{weight}. When
the energy scale is changed from $\mu_1$ to $\mu_2$ a mass $m_1$
will change to the value $m_2$, defining a functional relationship
$m_2 = m_2(m_1)$ or the reverse $m_1 = m_1(m_2)$. In order to
preserve the definition of the weight, it must transform as
\begin{equation}
\rho_{\mu_2 }(m_2) = \rho_{\mu_1} (m_1(m_2)) J(m_2)
\end{equation}
with
\begin{equation}
J=\frac{\partial m_1}{\partial m_2}.
\end{equation}
As an example, under QCD the masses transform as
\begin{equation}
m(\mu_2) = m(\mu_1) \left[\frac{\alpha_s (\mu_2)}{\alpha_s (\mu_1)}
\right]^{d_m}
\end{equation}
with
\begin{equation}
d_m = \frac{4}{11 -\frac{2N_f}{3}}.
\end{equation}
In this case the renormalization group transformation rule yields
\begin{equation}\label{gauge}
\rho_{\mu_2}(m) = \rho_{\mu_1}\left(m \left[ \frac{\alpha_s
(\mu_2)}{\alpha_s (\mu_1)}\right]^{-d_m} \right)
\left[\frac{\alpha_s (\mu_2)}{\alpha_s (\mu_1)} \right]^{-d_m}.
\end{equation}

For large values of the masses we also need to consider the effect
of the Yukawa interaction on its own running. Recall that in the
Standard Model the masses $m_i$ are related to the Yukawa couplings
$h_i$ via $m_i = h_i \, \frac {v} {\sqrt 2}$ where $v$ is the Higgs
vev. The renormalization group equations for the combined QCD gauge
and Yukawa running for $N_f =6$ are
\begin{eqnarray}
\frac{dg_3^2}{dt} &=& \frac{7}{16\pi^2}g_3^4   \nonumber \\
\frac{dh^2}{dt} &=& h^2\left( \frac{1}{2\pi^2}g_3^2
-\frac{9}{32\pi^2} h^2 \right)
\end{eqnarray}
where $t = \log (\mu_1^2/\mu^2)$, using $\mu_1$ as the initial scale.
Note that the QCD interaction tends to make the Yukawa coupling
larger as one scales down in energy ($t$ positive), while the Yukawa
self interaction tends to make the Yukawa coupling smaller. One of
the consequences of this is the well known quasi-fixed-point
\cite{ross} at $m_* =220 $~GeV - all large Yukawa couplings will run
to a value close to this fixed point when scaled down from large
energy to the $W$ scale. The solution to the Yukawa RGE has the form
\cite{ross}
\begin{eqnarray}
h(t) &=& \frac{b(t) h(0)}{[1 + a(t) h^2(0)]^{1/2}}  \nonumber \\
b(t) &=& \left( \frac{\alpha_s(t)}{\alpha_s(0)}\right)^{4/7}
\nonumber \\
a(t) &=& \frac{9}{2g_3^2(0)}\left[ \left(
\frac{\alpha_s(t)}{\alpha_s(0)}\right)^{1/7} -1 \right].
\end{eqnarray}
From this we can extract the Jacobian for the weight
\begin{equation}
J(m) = \frac{b^2}{\left[b^2- 2a \frac{m^2}{v^2} \right]^{3/2}}.
\end{equation}
For small masses, this is equivalent to the gauge rescaling given
above. For leptons, we neglect the QED effects and only consider the
Yukawa interaction. In this case, the above formulas continue to
hold, but with the identification
\begin{eqnarray}
b(t) &=& 1
\nonumber \\
a(t) &=& \frac{9}{32\pi^2} \ t.
\end{eqnarray}

Let us look at a particular weight which plays an important role in
the rest of this paper. This is a weight that has the form $\rho(m)
\sim 1/m$, which we will call the {\it scale invariant } weight. To
normalize this form we need to limit the range of masses with an upper
and a lower cutoff, so that the complete form is
 \begin{equation} \label{scaleinv}
   \rho \left( m \right) = \frac {1} {\log \frac {m_+} {m_{-}}} \ \frac {1} {m} \ \Theta
    \left( m - m_{-} \right) \Theta \left( m_+ - m \right).
\end{equation}
This distribution is scale invariant in two senses. First, a probability distribution that goes as $dm / m$ is clearly invariant under a rescaling of
all masses $m\to \lambda m$. The endpoints are not invariant but they rescale in the obvious fashion. In addition, this weight is invariant under
renormalization group transformation of the gauge interactions changing from one scale $\mu_1$ to another scale $\mu$. This can be seen simply from
Eq. (\ref{gauge}).

However, the Yukawa interactions modify the shape of this weight. In
particular, if the weight takes the above form at a scale $\mu_1$,
then at a lower scale $\mu$ it will have the form
\begin{equation} \label{scaleinv2}
   \rho_\mu \left( m \right) =\left[\frac{1}{1-2a \frac{m^2}{b^2v^2}}\right] \frac {1} {\log \frac {m_+} {m_{-}}} \ \frac {1} {m} \ \Theta
    \left( m - \hat{m}_{-} \right) \Theta \left( \hat{m}_+ - m \right)
\end{equation}
where
\begin{equation}
\hat{m}_\pm = \frac{b m_\pm}{\left[1+2a
\frac{m_\pm^2}{v^2}\right]^{1/2}}
\end{equation}
are the rescaled endpoints.
\begin{figure}
\begin{center}
$\begin{array} {c@{\hspace{0.01 in}}c} \multicolumn{1}{l}{} &
\multicolumn{1}{l}{} \\
{\resizebox{3in}{!}{\includegraphics{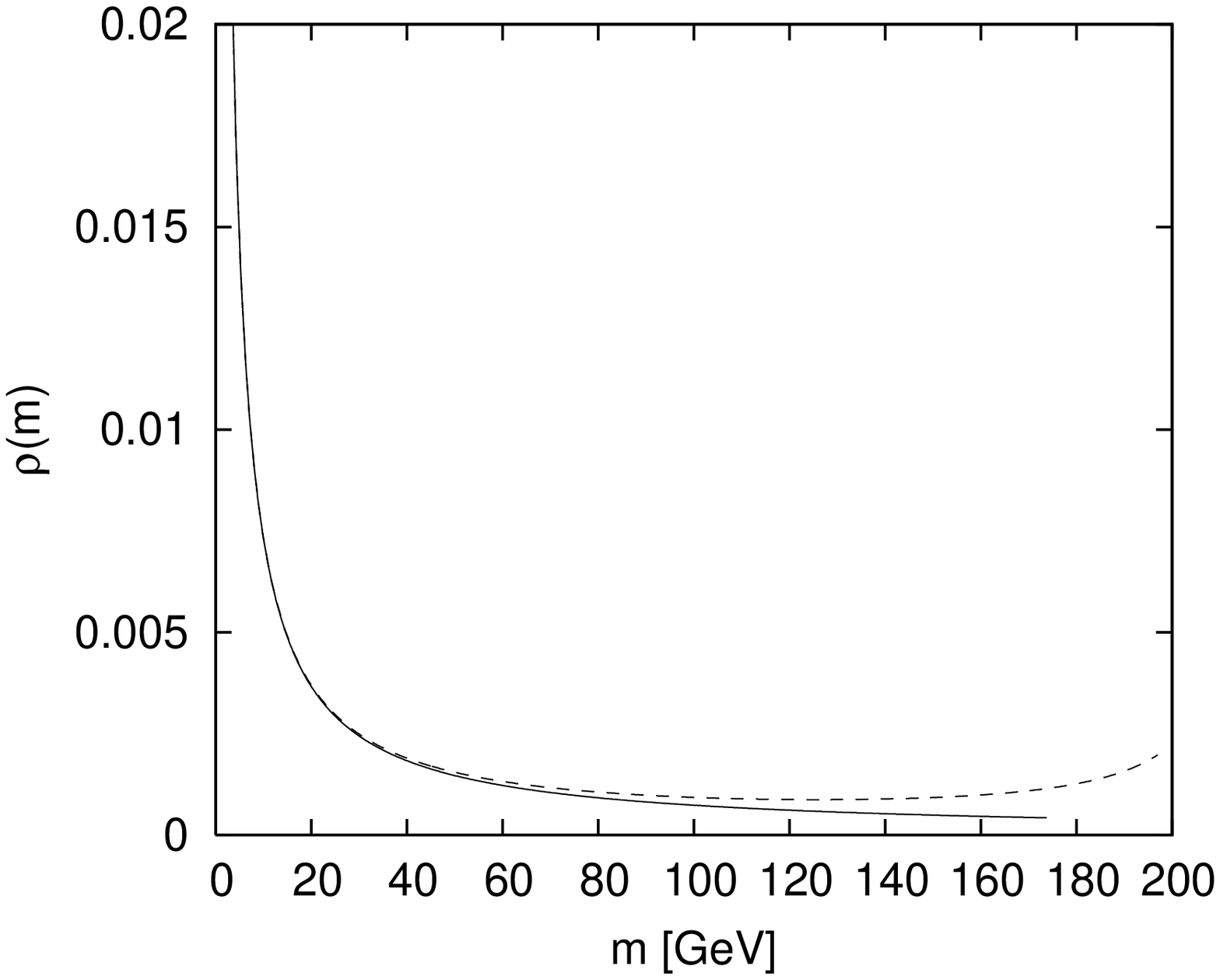}}}&
{\resizebox{3in}{!}{\includegraphics{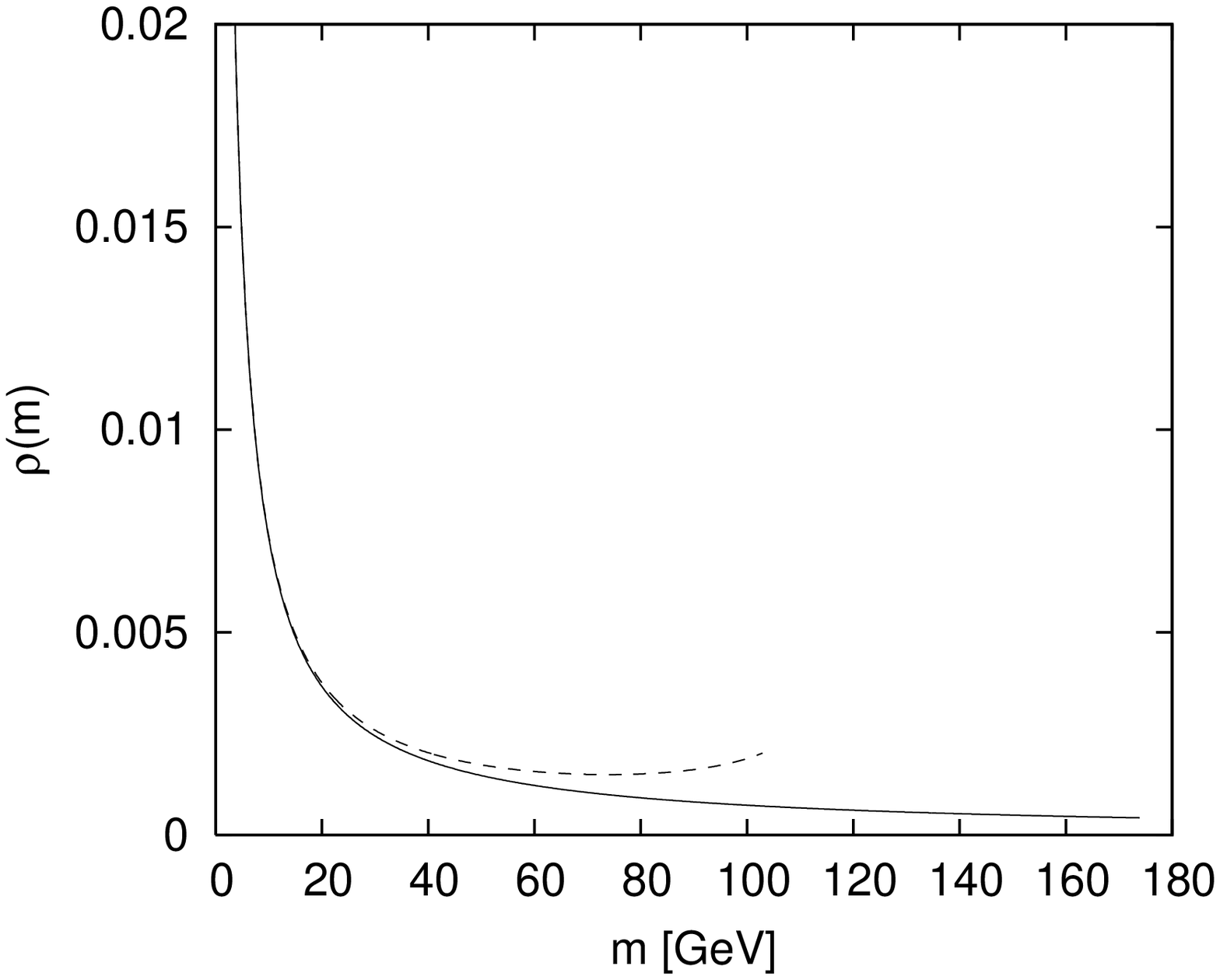}}} \\\\
\mbox{\hspace*{25pt} (a)} & \mbox{\hspace*{25pt} (b)}
\end{array}$
\end{center}
\caption{\small{(a) The solid curve corresponds to a pure scale invariant weight with an maximum Yukawa coupling of $1$ and a minimum of $1.2 \times
10^{-6}$. The dashed curve corresponds to using such a weight defined at the GUT scale and evolving it down to the $W$ scale using the
renormalization group equations for quarks. Note the slight distortion of the shape as well as the shift in the endpoints of the distribution.
(b) Same as before, except that the renormalization group for leptons was applied.}} \label{run}
\end{figure}

We can illustrate the renormalizaton group effects by considering
the transformation from a Grand Unified scale of $\mu = 10^{16}$~GeV
down to the $W$ scale. At the GUT scale we postulate for the Yukawa
couplings the weight
\begin{equation} \label{scaleinv3}
   \rho_\mu \left( h\right) = \frac {1} {\log \frac {h_+} {h_{-}}} \ \frac {1} {h} \ \Theta
    \left( h - {h}_{-} \right) \Theta \left( {h}_+ - h \right)
\end{equation}
with $h_+ = 1$ and $h_- = 1.2 \times 10^{-6}$. The values of the endpoints
have been chosen to allow the mass ranges to extend from below the electron
mass to above the top quark mass. When transformed down to the $W$ scale,
the maximum value of the quark mass distribution is $\hat{m}_+ = 197$~GeV
and the minimum value is $\hat{m}_- = 0.53$~MeV. For the leptons, the
corresponding values are $\hat{m}_+ = 103$~GeV and $\hat{m}_- = 0.20 $~MeV.
Note that the rescaled ranges of the distributions are different for quarks
and leptons. At low values of the mass the weight retains the $1/m$ form, but
the larger values the shape of the distribution is modified. This is shown in
Fig. \ref{run} (a) for the quarks and Fig. \ref{run} (b) for the leptons. The
modification to the shape is gentle enough that we will ignore it in most of the
phenomenological applications below. However, in the next section we
will briefly explore the effect of rescaling the quark and lepton masses
to the GUT scale and fitting the weight at that scale.

\section{Quark and charged lepton masses}

In this section we will treat the quark and charged lepton masses and will
determine the weight that best describes their distribution. We
describe the masses at the scale of $M_W$. Running the quark masses
to $M_W$ the dominant contributions come from QCD running and
we can neglect electroweak running for our purposes. The quark
masses at the scale $M_W$ \cite{weight} are given in Table \ref{quarkmassesatmwtable}.
For leptons we neglect the small effects on the masses from
electroweak running so that the lepton masses we use at $M_W$
are simply the ones quoted by the PDG \cite{PDG} as given in Table \ref{leptonmassesatmwtable}.

\begin{table}[h]
 \begin{center}
  \begin{tabular}{|c|c|c|c|c|c|} \hline
   $m_u$ & $m_d$ & $m_s$ & $m_c$ & $m_b$ & $m_t$ \\ \hline
   $2.2 \text{ MeV}$ & $4.4 \text{ MeV}$ & $80 \text{ MeV}$ & $0.81 \text{ GeV}$ & $3.1 \text{ GeV}$ & $170 \text{ GeV}$\\ \hline
  \end{tabular}
  \caption{\small{Quark masses at the scale $M_W$.}}
  \label{quarkmassesatmwtable}
 \end{center}
\end{table}
\begin{table}[h]
 \begin{center}
  \begin{tabular}{|c|c|c|}  \hline
   $m_e$ & $m_\mu$ & $m_\tau$\\ \hline
   $0.511 \text{ MeV}$ & $0.106 \text{ GeV}$ & $1.78 \text{ GeV}$ \\ \hline
  \end{tabular}
  \caption{\small{Lepton masses at the scale $M_W$.}}
  \label{leptonmassesatmwtable}
 \end{center}
\end{table}

From these values, we can deduce several properties.
The masses span a range of more than five orders of magnitude,
from $m_e$ to $m_t$. Most obviously, the weight is not flat,
but it must be peaked towards low masses. It is also not likely to
be a Gaussian or an exponential, because such distributions would
have an exponentially small probability of accommodating the top
quark. The weight must have a peak at low mass, yet extend out to
high mass in order to explain the top quark.

We will explore the class of power law weights, and will see that
they are successful in describing the distribution in masses.
The power law weights have the form
\begin{equation}
   \rho \left( m \right) \sim \frac {1} {m^\delta}.
\end{equation}
Obviously they are not normalizable if the range includes all values
of $m$, so we need to limit the range of the values of masses $m$.
For \(\delta \leq 1\) we need an upper bound of the range of $m$ so
that the weight is normalizable. A natural choice for that is the
quasi fixed point of the SM, \(m_* = 220 \text{ GeV}\), which we
will use as an upper bound for masses for all values of \(\delta\).
For \(\delta \geq 1\) we also need a lower bound for the range of
$m$. In order to explain the existence of the electron but no
lighter fermions, a sensible value for the lower bound should be
smaller than $m_e$ but not by many orders of magnitude.
We choose \(m_{low} = 0.4 \cdot m_e\) \cite{weight} where
$m_e$ is the electron mass. Varying this value does not significantly
affect the conclusions presented in this section.
Using these bounds and normalizing the weight,
we obtain the correct expressions for power law weights:\\
A) for \(\delta = 1\):
  \begin{equation} \label{rhodeltaone}
   \rho \left( m \right) = \frac {1} {\log \frac {m_*} {m_{low}}} \ \frac {1} {m} \ \Theta
    \left( m - m_{low} \right) \Theta \left( m_* - m \right)
  \end{equation}\\
B) for \(\delta \neq 1\) with a lower bound \(m_{low}\):
  \begin{equation}
   \rho \left( m \right) = \frac {1-\delta} {m_*^{1-\delta} - m_{low}^{1-\delta}} \ \frac {1}
    {m^\delta} \ \Theta \left( m - m_{low} \right) \Theta \left( m_* - m \right)
  \end{equation}\\
C) for \(\delta < 1\) without a lower bound \(m_{low}\):
  \begin{equation}
   \rho \left( m \right) = \frac {1 - \delta} {m_*^{1-\delta}} \ \frac {1} {m^\delta} \ \Theta
    \left( m_* - m \right)
  \end{equation}

\begin{figure}[ht]
 \begin{center}
  \includegraphics[scale=0.75]{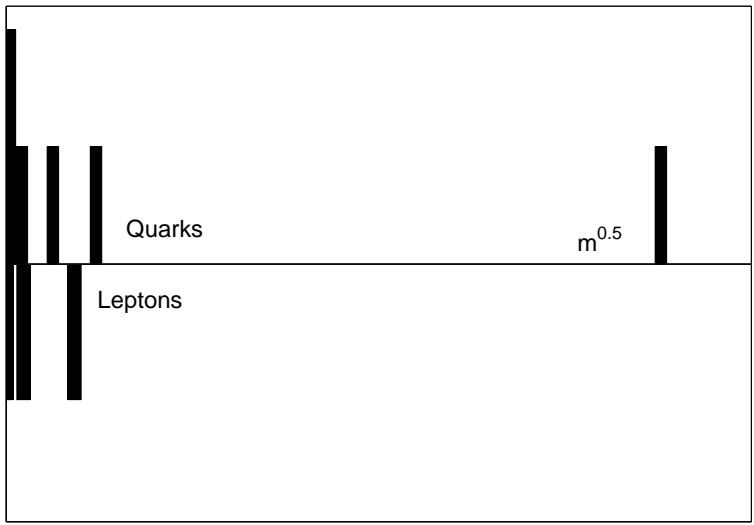}
 \end{center}
 \caption{\small{Quark and lepton masses, defined at the energy $\mu=M_W$, plotted versus $\sqrt{m}$. A
 weight $\ \rho \sim 1/\sqrt{m} $ corresponds to a uniform distribution on this scale.}}
 \label{rootm}
\end{figure}

The distribution of a sample of random numbers with a power law weight
 $\rho \left( m \right) \sim \frac {1} {m^\delta}$ looks like a uniformly
distributed sample on the scale $m^{1-\delta}$ (for $\delta \neq
1$). For the scale invariant weight $\rho \left( m \right) \sim
\frac {1} {m}$, a sample looks uniformly distributed on the scale
$\log m$. Fig. \ref{scaleinvariant} shows the physical masses of the
quarks and leptons on a log scale and it is visible that these are
consistent with a uniform distribution. For contrast, in Fig.
\ref{rootm} we plot the quark and lepton masses on a $\sqrt m$ scale
which corresponds to a weight $\rho \left( m \right) \sim \frac {1}
{\sqrt m}$. It is evident that this sample is quite unlikely to
result from a uniform random distribution.

We can make this mathematically precise by using the maximum
likelihood method to select the optimum power $\delta$. The
likelihood function is defined as the product of the weights of the
masses:
\begin{alignat}{2}
 L(\delta) & = \prod_{i=u,d,s,c,b,t} \rho \left(m_i,\delta\right) & \text{for quarks only} \ \ \ \ {}\\
 L(\delta) & = \prod_{i=e,\mu,\tau,u,d,s,c,b,t} \rho \left(m_i,\delta\right) & {} \ \ \ \ \ \ \ \text{for quarks and leptons}
\end{alignat}
The optimum power delta is where $L(\delta)$ has a
maximum $L_{max}$, which coincides with the maximum of the
log likelihood function $\log L(\delta)$. The one sigma
range is limited by the values of delta where
$\log L (\delta) = \log L_{max} - \frac {1} {2}$ and the two sigma
range is limited by the values of delta where
$\log L (\delta) = \log L_{max} - 2$.

\begin{figure}[ht]
 \begin{center}
  \includegraphics[scale=0.75]{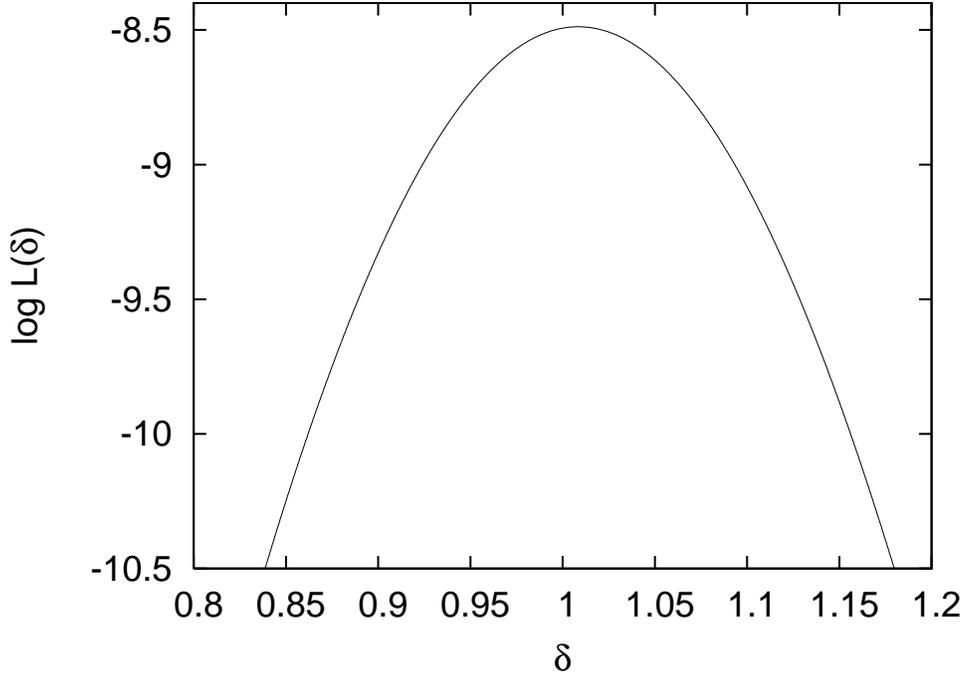}
 \end{center}
 \caption{\small{Log likelihood function for quark and lepton masses combined with $m_{low}$.}}
 \label{logli}
\end{figure}

We do this first in the range of masses where there exists a lower cutoff $m_{low}$
which we take at 0.4 of the mass of the electron. In this case all powers
of $\delta$ are allowed in principle. We perform the analysis for only the quark
masses and for the quark and lepton masses combined. The log
likelihood function $\log L \left( \delta \right)$ for the quarks and leptons
combined is plotted in Fig. \ref{logli}. In this case the favored
value of $\delta$ is almost exactly unity, with a range
\begin{equation} \delta = 1.02
\pm 0.08.
\end{equation}
 If we only use the
information in the quark masses we find nearly the identical result,
$\delta = 0.99 \pm 0.10$. The error bars in these results include
the effects of the limited statistics and get smaller when we
include the lepton masses.

\begin{figure}[h!]
\begin{center}
   \includegraphics[scale=0.35]{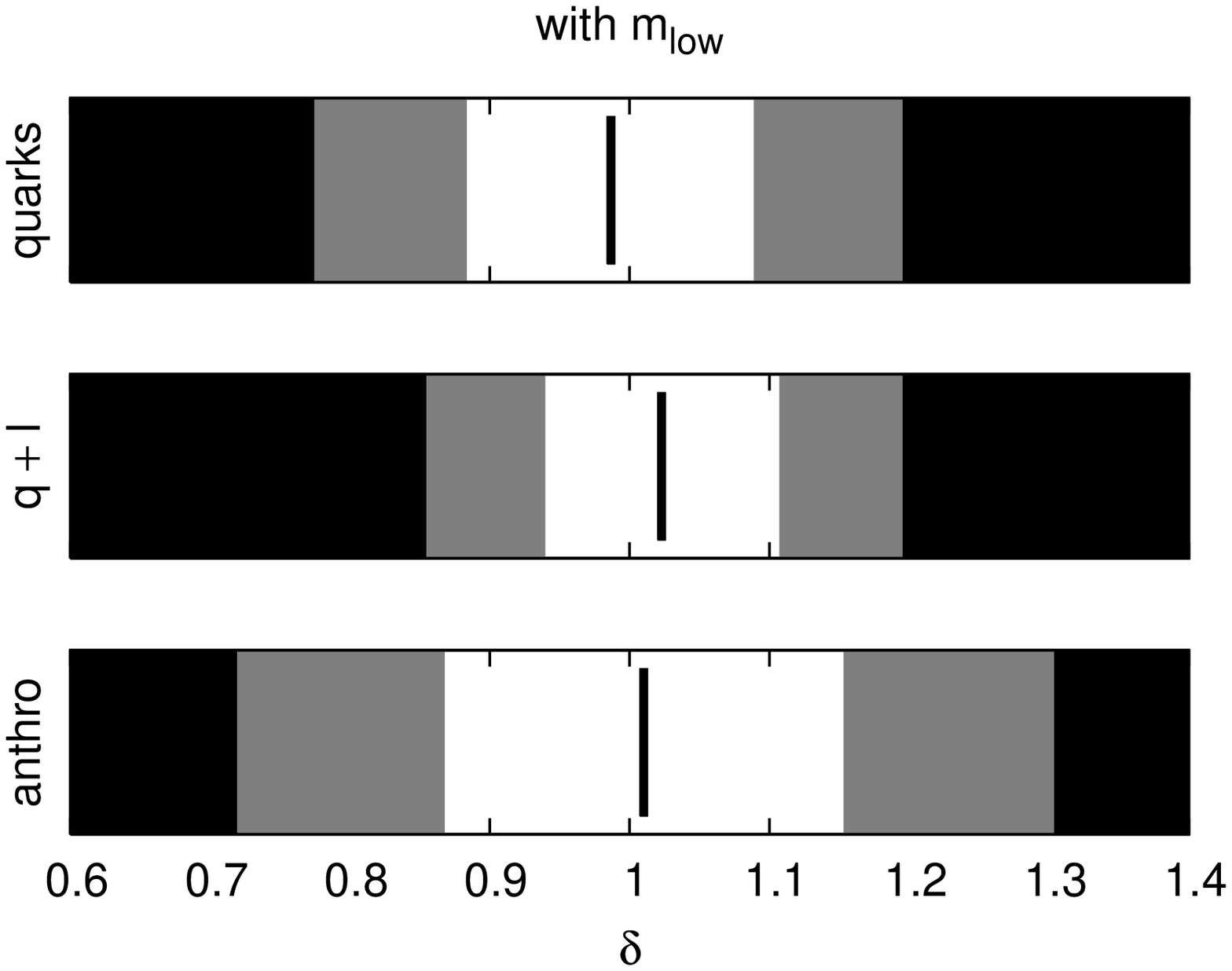} \ \ \ \ \
   \includegraphics[scale=0.35]{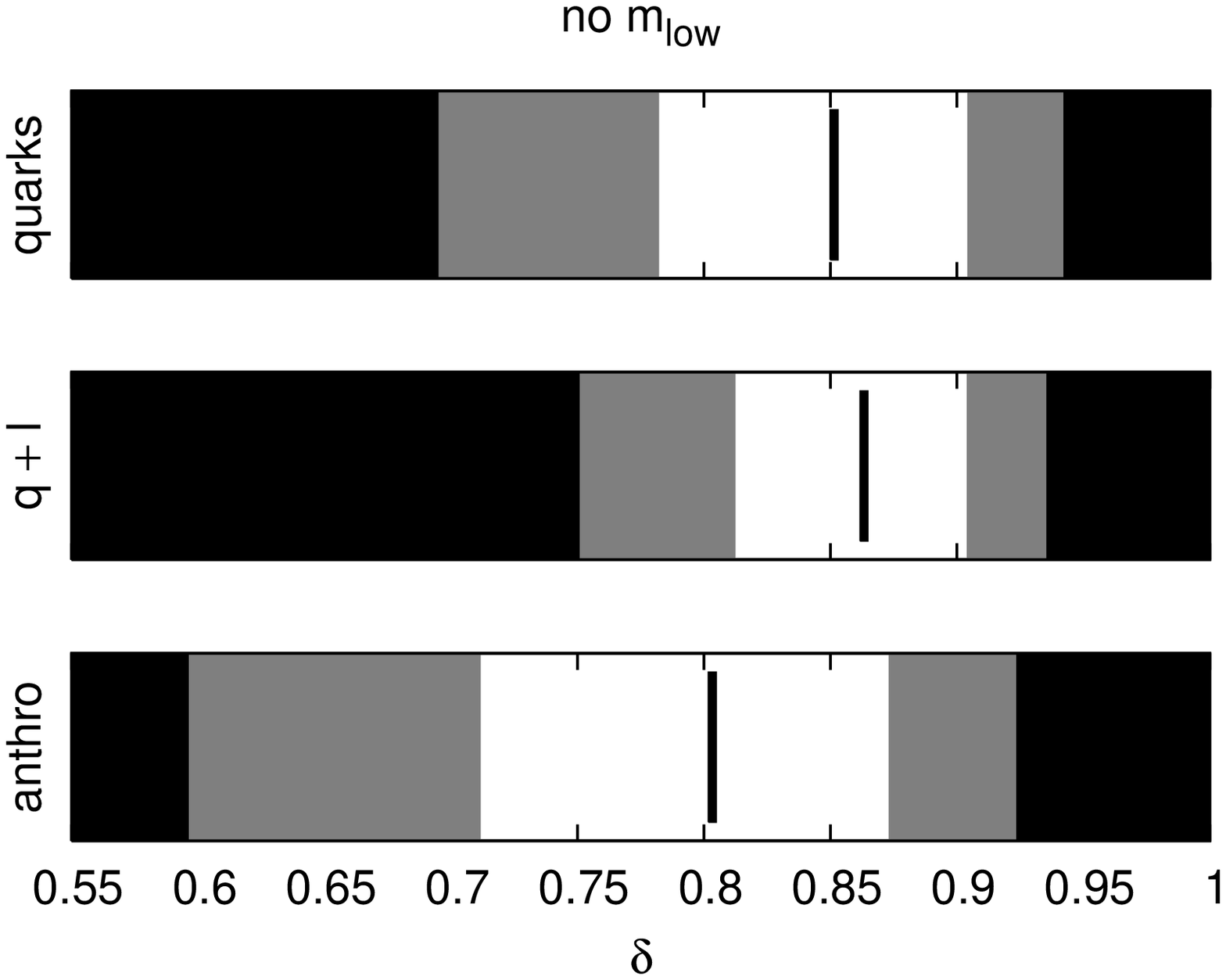}
   \end{center}
  \caption{\small{Results of the likelihood method. The black ranges are
           $2 \sigma$ excluded, the grey ones are $1 \sigma$ excluded
           and the lines inside the white regions are at the optimum
           powers $\delta$. The upper results are the ones taking
           only the quark masses into account, the ones in the middle
           are from quarks and leptons combined and the results at the
           bottom are the ones for quarks and leptons with anthropic
           constraints.}}
  \label{logliranges}
\end{figure}

Alternatively we could consider power law weights that extend all
the way down to $m=0$. In this case the power $\delta$ must be
strictly less than unity. We obtain $\delta =0.86^{+0.04}_{-0.05}$
for quarks and leptons and $\delta =0.85^{+0.05}_{-0.07}$ for quarks
only. All results from our likelihood fits are graphically
summarized in Fig. \ref{logliranges}.

\begin{figure}[h]
 \begin{center}
  \includegraphics[scale=0.75]{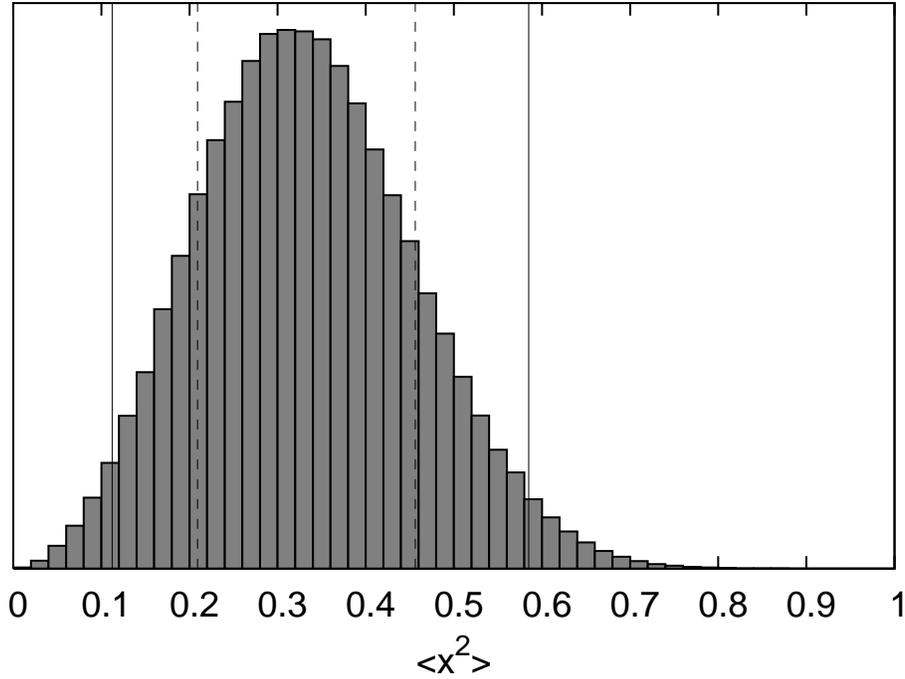}
 \end{center}
 \caption{\small{Distribution of $\big < x^2 \big>$ for six random variables. The dashed lines mark the $1 \sigma$ ranges and the solid lines mark the $2 \sigma$ ranges.}}
 \label{distrixsquared}
\end{figure}

Another way to probe the character of the weight is by a moment
analysis. A random variable x which is uniformly distributed
between 0 and 1 has moments
\begin{equation}
\big <x^n \big> = \frac{1}{n+1}.
\end{equation}
Of course for a small number of values drawn from a random
distribution the moments will not be precisely these values, but
will exhibit a scatter around these values due to the limited
statistics. We can understand this effect by simulations. As an
example we consider the second moment $\big < x^2 \big>$ of six
random variables drawn from a uniform distribution from zero to one.
We simulate six random numbers, evaluate $\big < x^2 \big>$ and
repeat that many times. That results in a distribution of $\big <
x^2 \big>$ which has the shape shown in Fig. \ref{distrixsquared}.
From this figure one can not only verify the expected average, but
also understand the 1$\sigma$ and 2$\sigma$ ranges. For example, at
two standard deviations we find the allowed range of the second
moment of six randomly selected numbers is
\begin{equation}
 0.112 \le \big < x^2 \big > \le 0.585.
\end{equation}

\begin{figure}[h]
\begin{center}
   \includegraphics[scale=0.35]{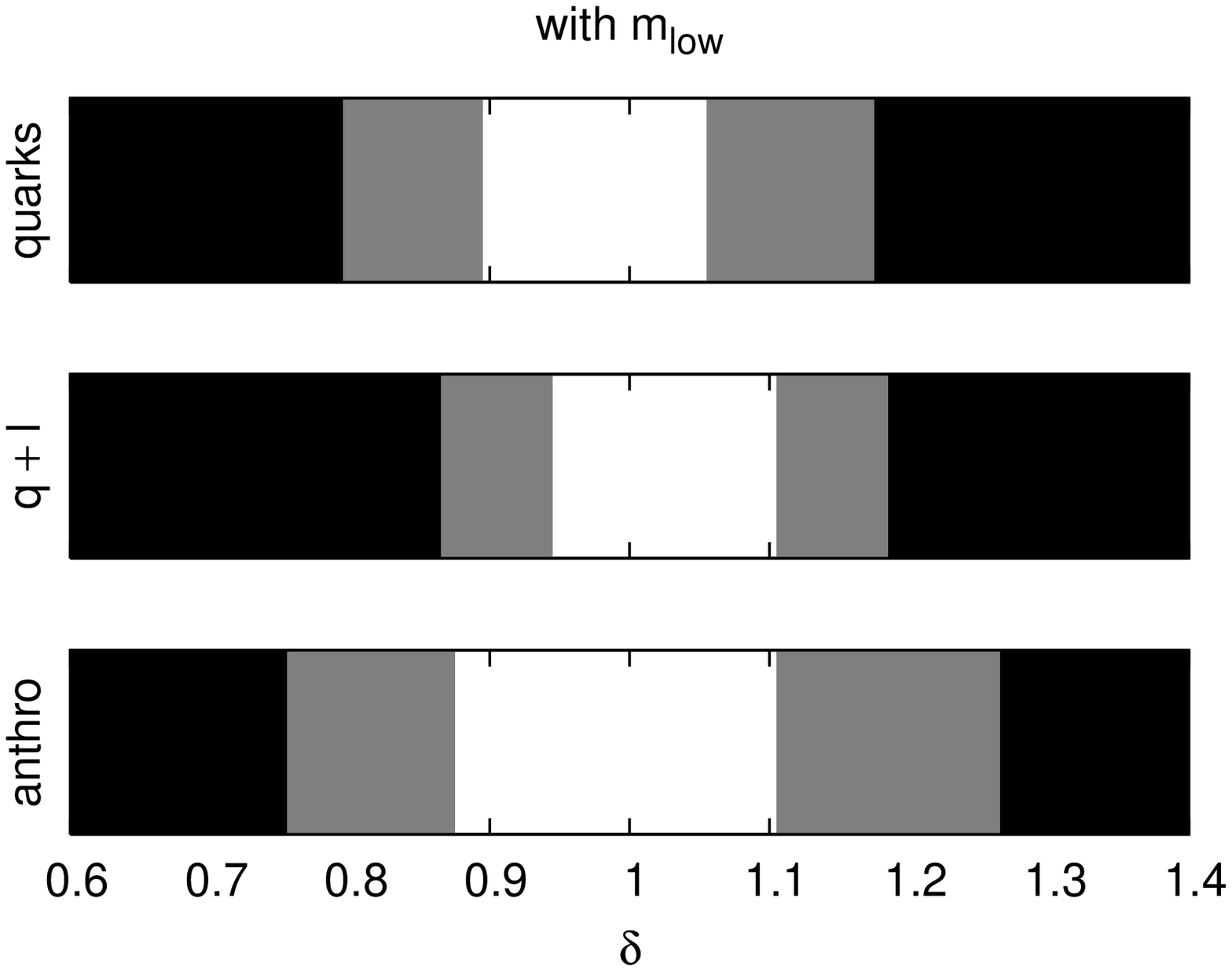} \ \ \ \ \
   \includegraphics[scale=0.35]{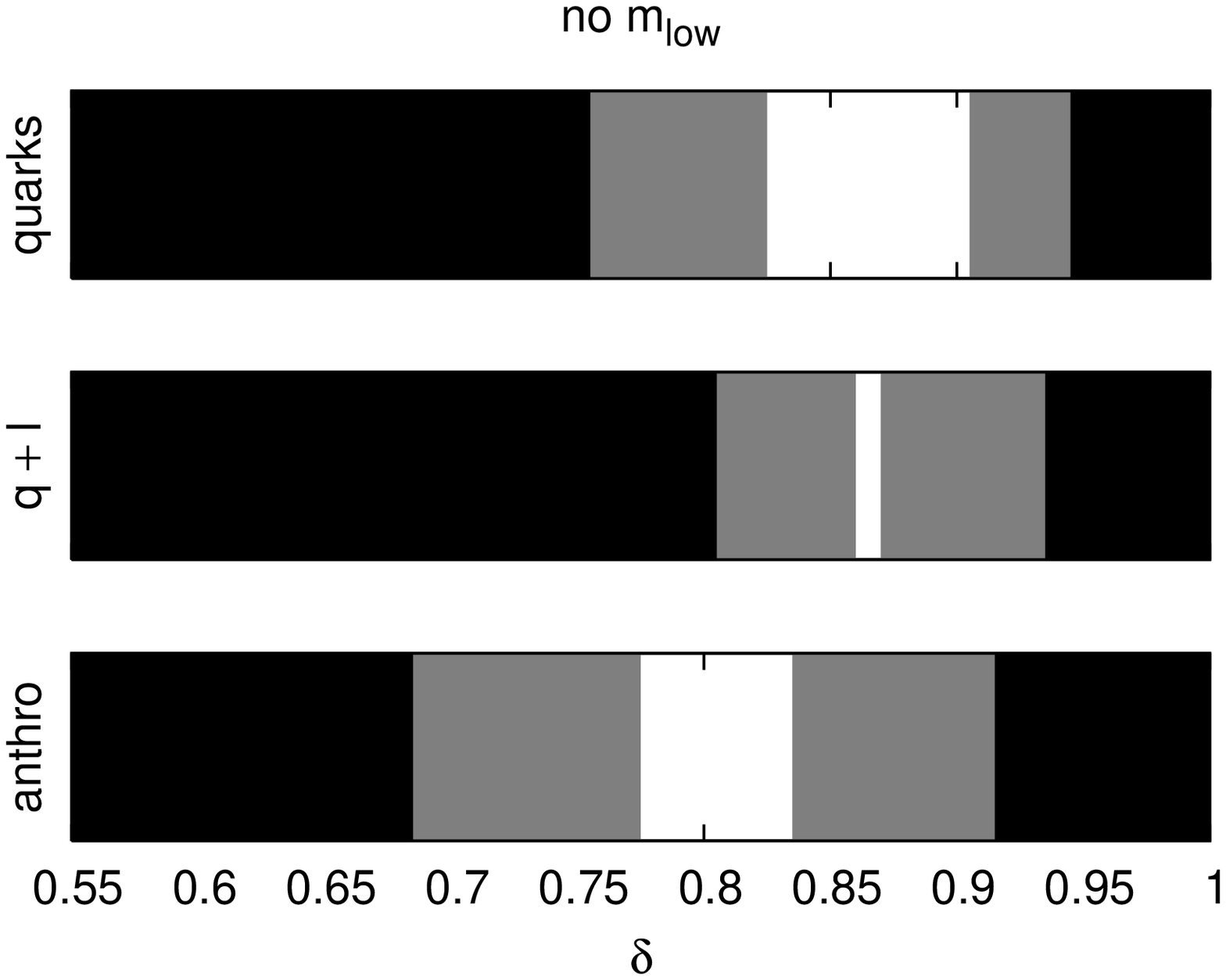}
   \end{center}
  \caption{\small{Results of the moment analysis. The black ranges are
           $2 \sigma$ excluded and the grey ones are $1 \sigma$ excluded.
           The upper results are the ones taking
           only the quark masses into account, the ones in the middle
           are from quarks and leptons combined and the results at the
           bottom are the ones for quarks and leptons with anthropic
           constraints.}}
  \label{momentsranges}
\end{figure}

We apply these tests to the weights, asking if the masses at $M_W$
correspond to a uniformly distributed random variable in
$m^{1-\delta}$ or $\log m$.
We use the first three moments $n=1, 2, 3$ and require that all of
these moments fall within the $1 \sigma$ or $2 \sigma$ ranges.
Besides the moments $\big < x^n \big >$, we also take into account
$\big <\left(1-x\right)^n \big >$ and
$\big <\left(\frac {1} {2} - x\right)^n \big >$ for $n=1, 2, 3$
and always take the most stringent bounds of all of them.
The results from this moment analysis are plotted in Fig.
\ref{momentsranges}.

Comparing with the $1 \sigma$ and $2 \sigma$
ranges from the likelihood method, we find that the allowed ranges
are somewhat smaller from the moments analysis but otherwise are
in good agreement. The fact that we always take the most stringent
bounds in the moment analysis probably lets the allowed ranges
shrink in comparison to the ones from the likelihood method.

One subtlety we did not discuss so far is the distortion of our
analysis by possible anthropic constraints. This is a difficult
subject and we cannot fully address it here. However, anthropic
issues have the potential to undermine our fundamental assumption
that the masses which we see are representative of the Standard
Model ensemble. This is because the formation of complex elements
which are necessary for life can only occur for certain
configurations of the charged fermion masses. In particular some
masses need to be light \cite{anthropic, agrawal, thibault} on the
QCD scale - otherwise the nucleons would decay rather than be bound
in the nucleus. Thus the masses that we see may represent an
anthropic bias in favor of light masses.

We can partially address anthropic issues by the following strategy.
We will eliminate from consideration the up and down quarks and the
electron, and will revisit our analysis only considering the second
and third generations of fermions. This is because the first
generation masses may be biased by anthropic considerations while
there are no known anthropic constraints on the heavier fermions.

Specifically, we proceed as follows. We drop the light masses $m_u$,
$m_d$ and $m_e$ from our analysis. We also exclude the possibility
that more than these three fermions are very light. This is not a
firm requirement as there could be different pattern of elements if
more quarks or leptons were light. However, this assumption that
only the three members of the first generation play a role in the
elements does correspond to the anthropic setting that we find
ourselves in, so it makes sense to study only this situation. We
apply this constraint by not allowing any other masses to occur
below $10$~MeV. This boundary is chosen because of the nuclear
binding energy of $10$~MeV/nucleon.  Fermions with masses greater
than this do not play a role in nuclear binding because they decay
weakly rather than being stable in the nucleus. For distributions
with a low mass cutoff, we accomplish this by taking $m_{low} = 10
\text{ MeV}$.

The results from our likelihood and moment analysis with anthropic considerations
are included in Fig. \ref{logliranges} and \ref{momentsranges}.
It is interesting that this modification does not significantly change our results.
For weights with a lower bound $m_{low}$ on the masses all
results are consistent with a scale invariant weight $\delta = 1$. For the weights
extending all the way down to $m=0$, a power $\delta$ in the range
$\delta = 0.7 - 0.9$ seems preferred. One can understand why there is little change
in our results by looking at the visual representation of the
distribution in Fig. \ref{scaleinvariant}. Our procedure for simulating anthropic
constraints consists of dropping the three lowest masses and applying a cutoff at 10
MeV, which on this figure is just above the mass of the down quark. One can see that the
remaining 4 quark and 2 lepton masses are consistent with
uniform distribution by themselves. Thus we have evidence for the scale invariant form
even considering only the heavier masses for which there is no
known anthropic constraints.

Finally, let us now consider the effect of scaling the quark and
lepton masses up to the GUT scale (taken to be $10^{16}$~GeV). We
could equivalently scale down a distribution from the GUT scale and
apply separate weights for quarks and leptons, as illustrated in
Section 2. However, it is simpler to use the renormalization group to
scale the masses up and study a common distribution at the GUT
scale. This explores the issue of whether the quarks and leptons
should be combined using the same distribution at the $W$ scale or
at the GUT scale. We will see that there is very little difference
between these alternatives. The renormalization group equations were
given in the previous section. We use only the Standard Model
interactions in the running of the masses up to the GUT scale. The
resulting Yukawa couplings defined at $10^{16}$~GeV are given in
Table \ref{YukawaGUTtable}.

\begin{table}[h]
 \begin{center}
  \begin{tabular}{|c|c|c|c|c|c|c|c|c|} \hline
   $h_u$ & $h_d$ & $h_s$ & $h_c$ & $h_b$ & $h_t$\\ \hline
   $0.46 \times 10^{-5} $ & $0.92 \times 10^{-5}$ & $ 1.7\times 10^{-4} $ & $1.7 \times 10^{-3} $
   & $6.5\times 10^{-3} $ & $0.63 $\\ \hline
  \end{tabular}
  \begin{tabular}{|c|c|c|} \hline
    $h_e$ & $h_\mu$ & $t_\tau$\\ \hline
    $2.9 \times 10^{-6}$ & $ 6.1 \times 10^{-4}$ & $0.01$\\ \hline
  \end{tabular}
  \caption{\small{Quark and lepton Yukawa couplings at the scale $M_{GUT}= 10^{16}$~GeV.}}
  \label{YukawaGUTtable}
 \end{center}
\end{table}

\begin{figure}[ht]
 \begin{center}
  \includegraphics[scale=0.75]{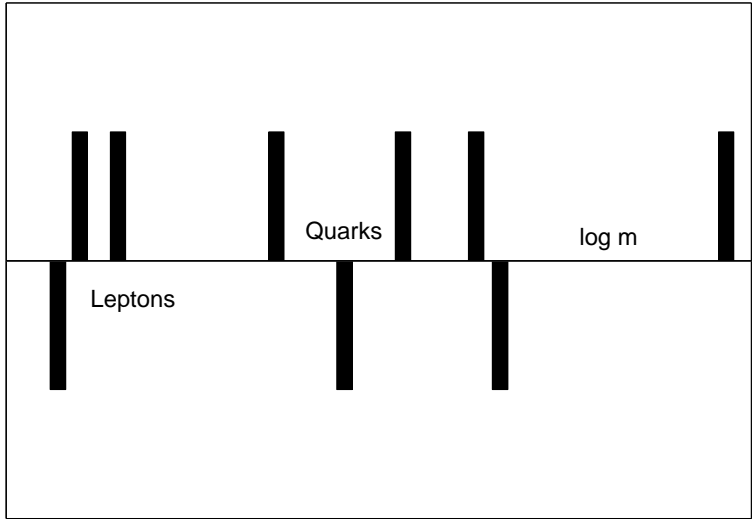}
 \end{center}
 \caption{\small{Quark and lepton Yukawa couplings, defined at the GUT scale
 $\mu=10^{16}$~GeV, on a logarithmic scale. This differs from Fig. \ref{scaleinvariant} by the
 (different) renormalization of the quark and lepton couplings as one transforms them to the
 GUT scale.}}
 \label{scaleinvariantGUT}
\end{figure}

Again we can visually explore the the comparison of these results
with the scale invariant weight by displaying them on a log scale,
as in Fig. \ref{scaleinvariantGUT}.
While there is a small shift in the relative placement of the quarks
and leptons, due to their different running up to the GUT scale,
this make no qualitative difference in the apparent randomness of
the distribution. This is also manifest in a maximum likelihood fit.
We consider the weights with both upper and lower cutoffs - in
particular we use $h_+ =1$ and $h_- =1.2 \times 10^{-6}$. As noted
previously, these correspond to a maximum quark mass of $197$~GeV
and a minimum lepton mass of $0.20 $~MeV. A maximum likelihood
analysis of the weight for the combined quark and lepton Yukawa
coupling as a function of $\delta$ again favors values of $\delta$
near one with a fit value of $\delta = 1.06 \pm 0.08$. The
$2\sigma$ allowed range of $\delta$ is $0.89-1.24$. These
ranges are very similar to the ones that we obtained when we simply
combined quark and lepton masses at the $W$ scale. We conclude from
this that effects of different treatments of quarks and leptons are
small. Also we have learned that it does not make any practical
difference if we determine the weight at the $W$ scale or at the GUT
scale - the distribution is close to scale invariant at either
energy.

\section{Quark mixing}

In this section we extend our discussion to the full Yukawa matrices
for up- and down-type quarks, generating not only the quark masses
but also the CKM mixing elements. Again, the assumption is that each
of the Yukawa elements can be treated as an independent variable to
be generated independently from a probability distribution function.
Of course, even in the landscape, this assumption could be incorrect
if there are symmetries or dynamics that constrain the independent
variables in the Yukawa matrices. However, we will see that this
ansatz does statistically account for the hierarchy that we see in
the CKM elements. The reason this occurs is simple - because the
weight is peaked at low mass, many of the Yukawa couplings are small
and therefore generate small mixing angles. This dominance of small
Yukawa elements may also provide an approximate explanation for the
idea of ``textures'' in the quark mass matrices~\cite{textures}.

The generation of the quark mass matrices is a well-known procedure
in the Standard Model \cite{dynamics}. The mass matrix for the
up-type quarks, for example, follows from the Yukawa coupling to the
Higgs through the relation
\begin{equation}
M^{(u)}_{0ij} = \frac{h_{ij}}{\sqrt{2}} v
\end{equation}
where $h_{ij}$ are the Yukawa couplings. This consists of 9
independent complex elements. The diagonalization of this matrix
occurs through separate transformations on the left handed and right
handed quarks
\begin{equation}
     m^{(u)} = \left(
      \begin{array}
       {*{3}{c}}
       m_u & 0 & 0 \\
       0 & m_c & 0 \\
       0 & 0 & m_t \\
      \end{array} \right) = V_L^{(u) \dagger} M_0^{(u)} V_R^{(u)}.
\end{equation}
A similar process occurs for the down-type quarks
\begin{equation}
     m^{(d)} = \left(
      \begin{array}
       {*{3}{c}}
       m_d & 0 & 0 \\
       0 & m_s & 0 \\
       0 & 0 & m_b \\
      \end{array} \right) = V_L^{(d) \dagger} M_0^{(d)} V_R^{(d)}.
\end{equation}
The CKM matrix is then formed by the product of the two left handed
rotation matrices
    \begin{equation}
     V_{CKM}=V_L^{(u) \dagger} V_L^{(d)}.
    \end{equation}

In general the elements of the mass matrix $M_{0ij}$ are
complex-valued. In our model the magnitude of the elements of
$M_{0ij}$ are being distributed from $m_{low}$ to $m_{*}$ according
to the weight $1/m^{\delta}$. We take the phases of
the elements of $M_{0ij}$ to be uniformly distributed between $0$
and $2\pi$. The appearance of complex elements for $M_{0ij}$ is
required for the existence of CP violation and of course is not
forbidden by any of the symmetries of the Standard Model. However,
we have also simulated the mixing matrices with real elements and we
note that this does not change any of our qualitative results, aside
of course from the measure of CP violation.

A first question to be addressed is whether an assumed weight for
the Yukawa couplings will lead to the same weight being applicable
for the masses that emerge as output from the diagonalization
process. We cannot address this question analytically, so we
approach it via simulations. The result that we find is that there
is a change in the weight between the initial Yukawa couplings and
the final masses and that this change is roughly categorized by a
shift in the power $\delta$ by about $-0.16$.
\begin{figure}
\begin{center}
$\begin{array} {c@{\hspace{0in}}c} \multicolumn{1}{l}{} &
\multicolumn{1}{l}{} \\
{\includegraphics[scale=0.4]{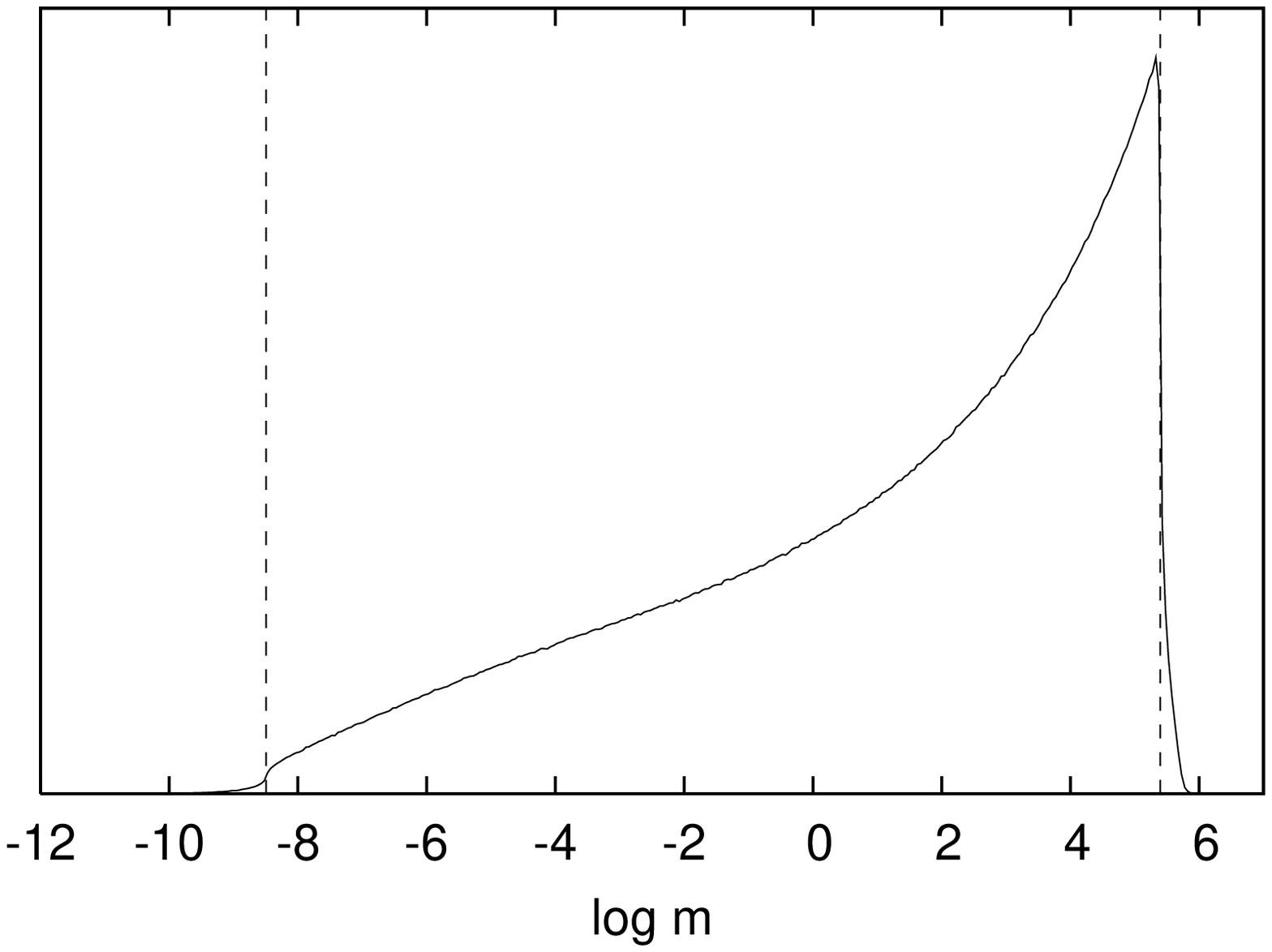}}&
{\includegraphics[scale=0.4]{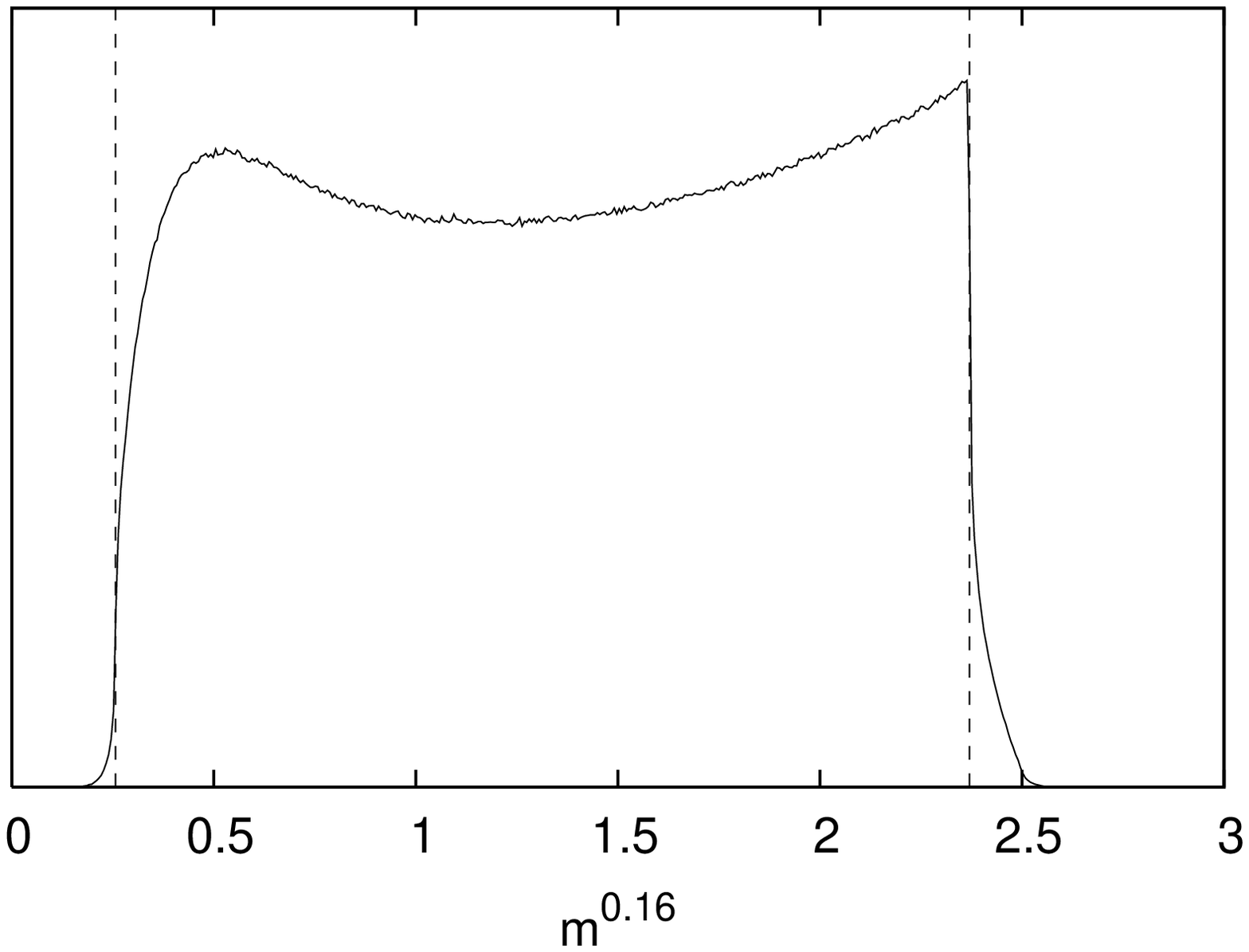}} \\\\
\mbox{(a)} & \mbox{(b)}
\end{array}$
\end{center}
\caption{\small{Elements of mass matrices have been distributed
according to the scale invariant weight. In $(a)$ we bin and plot the the
eigenvalues of the mass matrices on a logarithmic scale, whereas in
$(b)$ we do the same on an $m^{0.16}$ scale. The vertical dashed lines
indicate the values of $m_{low}$ and $m_{*}$. }} \label{eigenvalues}
\end{figure}

\begin{figure}[ht]
 \begin{center}
  \includegraphics[scale=0.75]{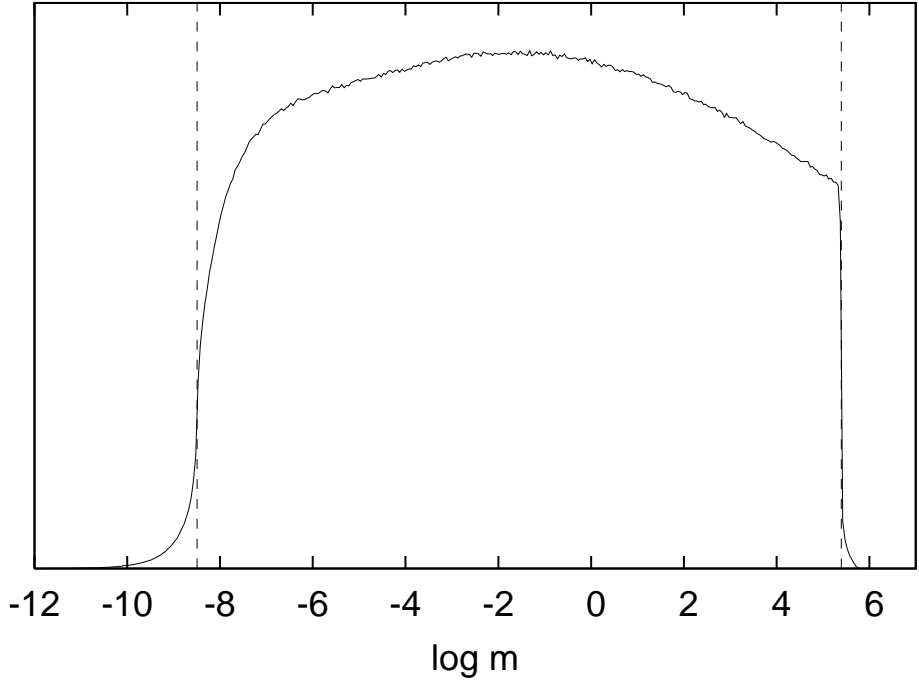}
 \end{center}
 \caption{\small{Distribution of the eigenvalues of the mass matrices, on
  a logarithmic scale, when the elements of the mass matrices
  have been distributed according to weight $\rho \sim 1/m^{1.16}$.}}
 \label{eigenvalues_delta_1.16_logscale}
\end{figure}

In order to demonstrate this, we have generated a large sample of
mass matrices of Yukawa couplings with a scale invariant weight
$\rho \sim 1/h$, diagonalized the matrices and bin and plot the resulting
mass values on a logarithmic scale in Fig. \ref{eigenvalues} (a). If
the resulting distribution was also scale invariant, this plot would be
flat. However, the result is not flat. However, when plotted versus
$m^{0.16}$ in Fig. \ref{eigenvalues} (b), we see that the resulting
distribution is almost flat, indicating that the result is
compatible with a weight of $\rho \sim 1/m^{0.84}$. Likewise if we
simulate a weight for the Yukawa interactions of $\rho \sim
1/h^{1.16}$ and bin and plot the resulting masses on a logarithmic
scale in Fig. \ref{eigenvalues_delta_1.16_logscale} we see an almost
flat distribution which would correspond to a scale invariant weight
for the output masses.

Another outcome we see is that after the diagonalization of the
Yukawa matrices the limits $m_{low}$ and $m_{*}$ are washed out for
the mass eigenvalues. There are masses smaller than $m_{low}$ and
larger than $m_{*}$, but as one can see in
Fig. \ref{eigenvalues} and \ref{eigenvalues_delta_1.16_logscale} masses which are orders
of magnitude smaller than $m_{low}$ or larger than $m_{*}$ are
highly suppressed.

A weight for the Yukawa matrix elements that is capable of explaining
the observed quark masses must therefore have a higher power $\delta$
than what we found in our analysis in Section 3. As we saw, $\delta = 1.16$
for the Yukawa matrix elements roughly corresponds to a scale invariant
distribution for the masses. The previous limits $m_{low}$ and $m_*$
can also be used for the Yukawa matrix elements.

\begin{figure}[ht]
 \begin{center}
  \includegraphics[scale=0.75]{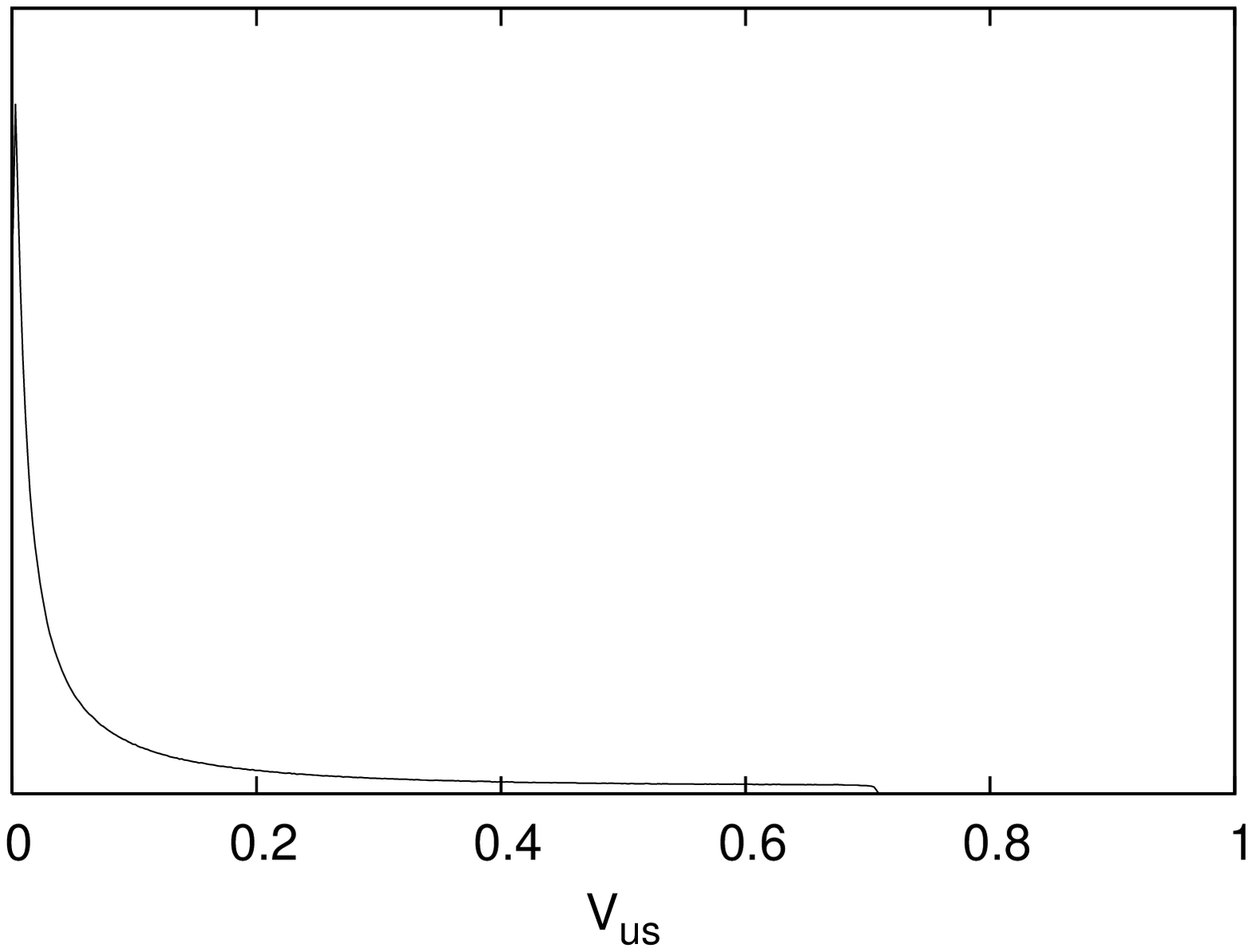}
 \end{center}
 \caption{Distribution of the simulated values of $V_{us}.$}
 \label{Vus}
\end{figure}

Now we turn to the details of the simulation of the CKM matrices.
First, we generate a large sample of mass matrices for both up-type
quarks and down-type quarks. Each element of a particular mass
matrix is taken to be independent of each other. By convention the
largest masses in the two mass matrices are the top and the
bottom, so we use the freedom to relabel the fields in order to
place the largest element in the (3,3) position of the mass
matrices.

As discussed above, the diagonalization of the mass matrices is
being done by biunitary transformations. In general, in this
procedure the physical mass eigenvalues of quarks have no constraint
to fall within any particular range. However for this mixing
analysis we wish to consider only output masses that are somewhat
similar to those observed in nature. We impose this by considering
only those mass matrices for which mass eigenvalues for up and down
quarks ($i.e$ the lightest eigenvalues of each matrix) are less than 10
MeV, mass eigenvalues for charm and strange quarks are within 50 MeV
and 2 GeV and mass eigenvalues for top and bottom quarks are greater
than 3 GeV.

We construct CKM elements from two left handed rotation matrices,
one from up-type quarks and the other from the down-type quarks. In
the results, there are a few discrete classes for the pattern of
mixing angles. Since we have not imposed the generation structure,
there is a set of cases where the up and the strange are dominantly
coupled to each other and $V_{ud}$ is small. Likewise there is a
small number of cases where the charm and bottom quarks are
dominantly coupled to each other. We discard these cases and only
consider those results which exhibit the proper generation structure
such that the largest couplings are to members of the same
generation (as defined by the relative masses).

We use a weight for the Yukawa couplings of $1/h^{1.16}$, which
generates the almost scale invariant weight for the masses. With
these constraints, the resulting distribution of the magnitude of
the $V_{us}$ elements of CKM matrices is shown in Fig. \ref{Vus}.
One can see that the distribution is peaked at lower values.
When one looks more closely at the distribution of $V_{us}$ at very
low values, one sees that the peak is located at about 0.002.\footnote{When
doing the same simulations without $m_{low}$ this peak does not
occur but the distribution for $V_{us}$ reaches a maximum at zero.
One can understand that from dimensional analysis: in the
simulations with $m_{low}$ there are two dimensionful parameters
involved, $m_{*}$ and $m_{low}$ where $m_{low} / m_{*} \ll 1$.
However, in the case without $m_{low}$ there is only one
dimensionful parameter. Therefore a peak for a dimensionless
quantity such as $V_{us}$ is expected to be found either at zero or
at a value of order one in a case where only one dimensionful
parameter exists.}
Another feature of the distribution is that it vanishes for
values larger than $1 / \sqrt 2$. The reason for that is the
generation structure we impose where we require the diagonal
elements to be the largest elements in their row and column in the
CMK matrix.

The other CKM elements we consider are $V_{ub}$ and $V_{cb}$. For
both we observe shapes of their distributions similar to the one for
$V_{us}$ as seen in Fig. \ref{Vus}, but with a higher preference for
small values.

For the distributions of the CKM elements one cannot define $1
\sigma$ and $2 \sigma$ ranges as for Gaussian distributions since
the peaks are very close to zero but the tails of the distributions
are extremely long. Therefore we instead consider the medians of the
distributions and the values where $68\%$ and $95\%$ of the elements
lie below them. We call the ranges up to these values the $1 \sigma$
and $2 \sigma$ ranges respectively. For $\delta = 1.16$ the observed
values of $V_{ub}$ and $V_{cb}$ fall within the $1\sigma$ ranges and
the physical value for $V_{us}$ is within the $2 \sigma$ range.

\begin{figure}[ht]
 \begin{center}
  \includegraphics[scale=0.75]{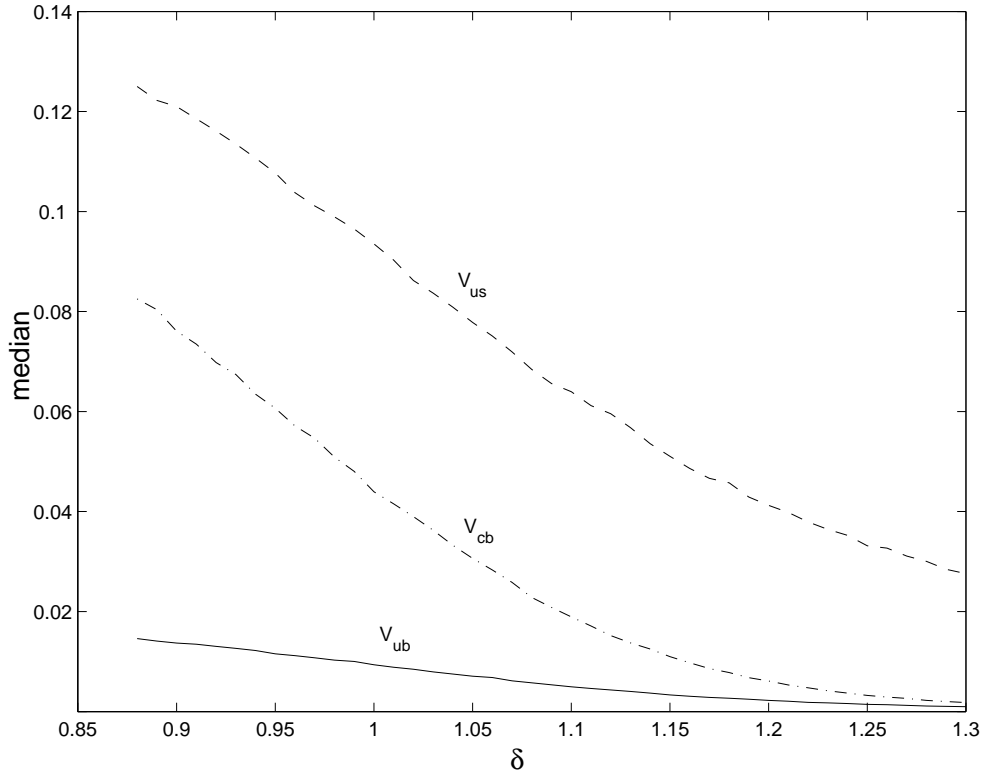}
 \end{center}
 \caption{Medians of the CKM elements as a function of $\delta$}. \label{medians}
\end{figure}

The medians of the CKM elements follow the similar hierarchy that is
observed in nature. In Fig. \ref{medians} we plot the medians of
$V_{ub}, V_{cb}, V_{us}$ as a function of $\delta$. In general,
there is a hierarchy in the magnitudes of the elements. This is
understandable from the diagonalization procedure. Because of the
$1/h^{\delta}$ distribution, most Yukawa couplings are small. The
mixing angles needed to diagonalize the mass matrices are
proportional to the off-diagonal elements divided by the difference
in masses. The mixing involving heavy top and bottom quarks will be
smaller than those that involve the mixing of down and strange
quarks. Thus we see that our model can accommodate the hierarchical
structure of CKM elements.

\begin{figure}[ht]
 \begin{center}
  \includegraphics[scale=0.75]{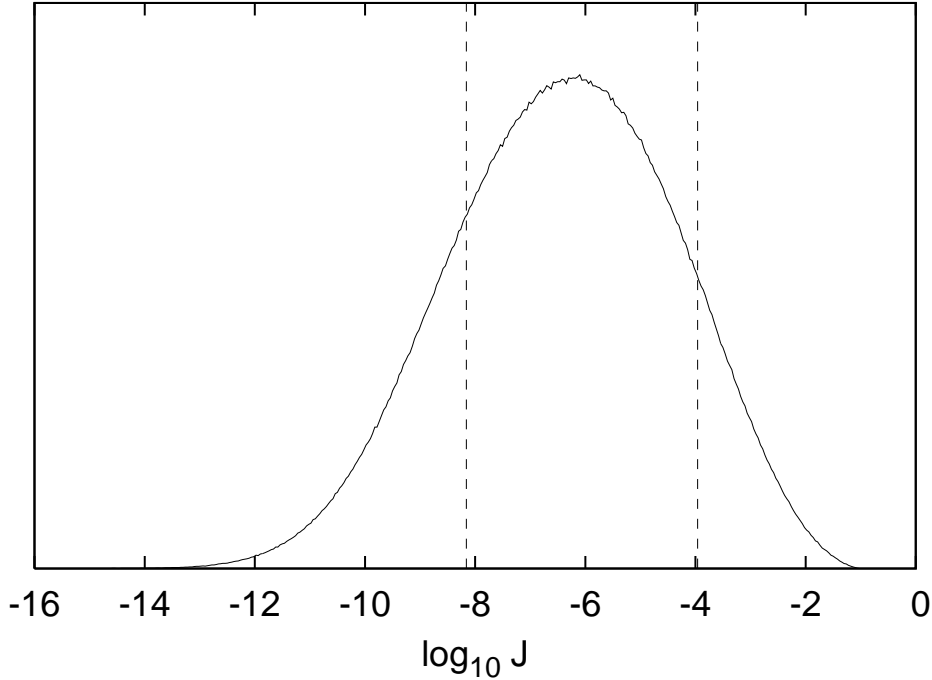}
 \end{center}
 \caption{The Jarlskog invariant for quarks describing the magnitude of CP violation}
\label{Jinvariant}
\end{figure}

After absorbing several complex phases into the definition of the
quark fields, the CKM matrix can be expressed by four parameters, of
which three are mixing angles and one is a complex phase. This
complex phase indicates the existence of CP violation in the theory.
Perhaps the best way to describe CP violation is through the
``Jarlskog invariant'', $J$ \cite{jarlskog} which is invariant under
rephasing of the quark fields. All CP violation is proportional to
the product of CKM elements
\begin{equation}
J = \text{Im}\left(V_{ud} V_{cs} V_{us}^* V_{cd}^*\right).
\end{equation}
In nature, this is observed to have the value $J = (2.88 \pm 0.33)
\times 10^{-5}$ \cite{PDG}. In Fig. \ref{Jinvariant} we plot the
distribution of $J$ and indicate the $1\sigma$ range. We note that
observed value of $J$ in nature is well within the $1\sigma$ range.
We conclude that, within the hypothesis of the weight, the observed
magnitude of CP violation is quite natural for the same probability
distributions that describe the quark masses.

\section{Neutrino masses and mixing}

The neutrino masses are different enough from those of the other
fermions that it is clear that they should not be governed by the
same weight. Moreover, there are strong theoretical reasons why
neutrino masses should be treated differently. In this section we
explore the description of the neutrino masses and mixing in the
case where they are described by the Type I seesaw mechanism
\cite{seesaw}. Specifically we will assume that the Dirac component
of neutrino masses is governed by the same weight as the charged
leptons, and will add the possibility of a different weight for the
Majorana component of the mass. Unfortunately our exploration will
be somewhat inconclusive. The observed pattern of masses and mixings
is not highly typical of the theoretical distributions, but it is
also not statistically excluded at the 2$\sigma$ level. We note that
other attempts at simulating neutrino properties with a statistical
distribution can be found in \cite{goldman} and \cite{anarchy}. We
compare our analysis to these at the end of this section.

Our present knowledge of neutrino masses comes from various neutrino
oscillation experiments and from the measurements of the anisotropies
in the CMB by WMAP and large scale structure formation.
From neutrino oscillation experiments, we know two mass differences
and WMAP gives an upper bound on the sum of the neutrino masses \cite{neutrino, wmap}:
\begin{equation*}
 5.4 \times 10^{-5} \text{ eV}^2 \leq \Delta m_{12}^2 \leq 9.5 \times 10^{-5} \text{ eV}^2
\end{equation*}
\begin{equation*}
 1.2 \times 10^{-3} \text{ eV}^2 \leq |\Delta m_{13}^2| \leq 4.8 \times 10^{-3} \text{ eV}^2 \\
\end{equation*}
\begin{equation}
 \sum_{k}m_{{k}} \leq 0.7 \text{ eV}
\end{equation}

First we consider the neutrino masses by themselves without mixing.
In the Type I seesaw mechanism, the light neutrino masses $m_i$ are
of the form
\begin{equation}
m_i = \frac{m_{D,i}^2}{M_M}
\end{equation}
where $m_{D,i}$ are the Dirac masses connecting left and right
handed fields and $M_M$ is the Majorana mass term of the right
handed neutrinos.

We assume that the Dirac masses come from the same weight as the
quark and charged lepton masses $\rho \sim 1 / m^\delta$ and that
the Majorana mass $M_M$ is a common scale. Since in the Type I
seesaw, the Dirac masses are proportional to the Higgs vev, it is a
reasonable assumption that the weight is the same as that for the
quarks and charged leptons. Note that a scale invariant weight for
the Dirac masses predicts that the neutrinos will also be
distributed with a scale invariant weight. If the Dirac masses are
distributed with a weight $\sim 1/m^\delta$, and the Majorana mass
is taken as a constant, then the neutrino masses are distributed
with a weight
\begin{equation}
\rho_{\nu}(m_\nu)\sim \frac{1}{m_\nu^{\delta_\nu}} \sim
\frac{1}{m_\nu^{2\delta-1}}
\end{equation}
and the use of $\delta =1$ implies $\delta_\nu =1$ also.

An initial result is that this scheme greatly favors a ``normal''
hierarchy in the neutrino masses with $m_1 \ll m_2 \ll m_3$. In this
case the solar mass difference is representative of the mass scale
of the second lightest neutrino with mass $m_2$. The third neutrino
is then heavier, with a mass $m_3$ of order the atmospheric mass
difference. The scale invariant weight favors this because it favors
small masses. It would be unlikely that there are two large masses
that just happened to be so close together to yield the small solar
mass difference. In the simulations that we describe below for the
case involving mixing, the probability of obtaining the mass
differences with an inverted hierarchy compared to a normal
hierarchy was about 0.1\%. A quasi degenerate neutrino mass spectrum
is even much more unlikely. From now on we will assume that a
``normal'' hierarchy exists.

One might hope that the observed masses, combined with the
assumption that the Dirac masses contribute to the seesaw mechanism
with the weight determined for the other fermions, would lead to a
prediction for the scale of the Majorana mass. In practice this does
not occur. We have performed a likelihood fit for the scale of the
Majorana mass $M_M$ and we find that the $2 \sigma$ allowed range
for $M_M$ spans from $10^5 \text{ GeV}$ to $10^{15} \text{ GeV}$.
The details of this fit are given in the Appendix.

\begin{figure}[ht]
 \begin{center}
  \includegraphics[scale=0.75]{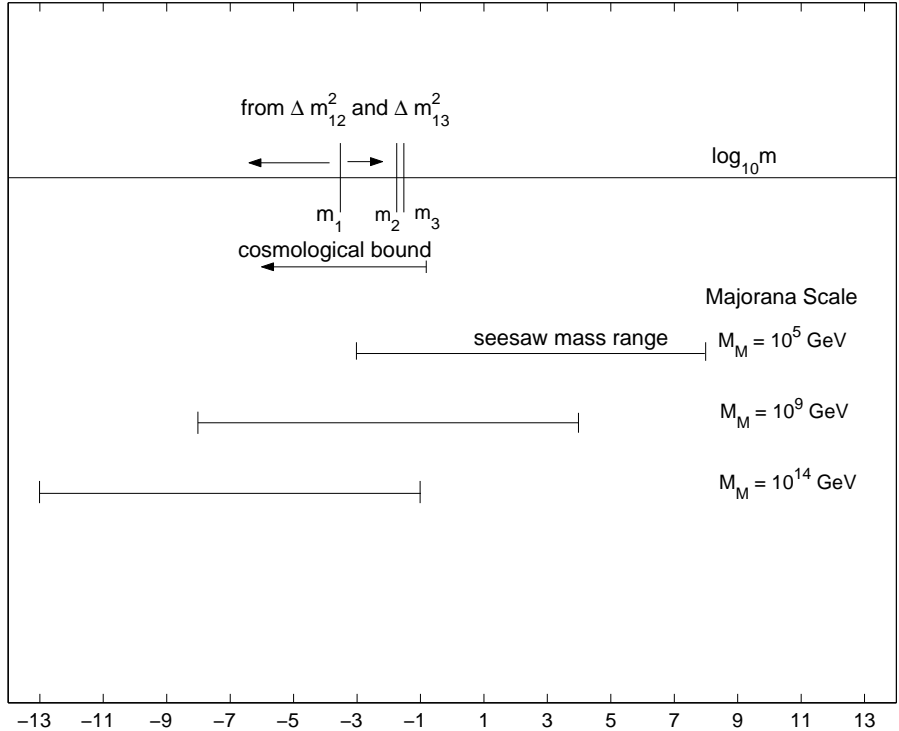}
 \end{center}
 \caption{The neutrino masses on a logarithmic scale. The mass differences
 were translated into rough mass estimates as described in the text. Note that the lightest
 mass is unconstrained at small values of the mass. If the distribution is to be
 described by a scale invariant weight, these masses should be uniformly distributed on
 this log scale. The range of the distribution is also shown for various values of the
 Majorana mass. Given the limited statistics, it is hard to determine if a scale invariant
 distribution is appropriate or to bound the Majorana mass scale.}
\label{neutrinos}
\end{figure}

One way to see why the limits on the Majorana mass scale are that weak
is to plot the masses on a log scale. Of course we
do not know the absolute masses precisely since we have only the
mass differences. However, under the assumption of a normal
hierarchy we know the largest mass quite well and, if the lightest
mass is not anomalously close to the second lightest mass, we have a
reasonable estimate of the second lightest mass. However, there is
no constraint at all on the mass of the lightest neutrino. Thus the
situation is close to that pictured in Fig. \ref{neutrinos}, where the
rough positions of the two heaviest neutrinos are pictured and the
lightest is allowed at almost any location smaller than these on the
logarithmic scale.

In the figure also are shown the ranges allowed for
the neutrino masses in the situation where the Dirac mass is
described by the scale invariant weight and the Majorana mass scale
$M_M$ is a constant. The ranges are shown for various values of
the Majorana mass scale $M_M$. The scale invariant distribution would
correspond to a uniform random distribution on this logarithmic scale.
It is clear visibly that if we allow the lightest
neutrino mass to be in a broad possible range, that a wide range of
Majorana masses is consistent with the scale invariant mass
distribution. For high Majorana scales $M_M$ there is however more
``phase space'' for the lightest mass which leads to the
slight preference for high Majorana mass scales $M_M$ as seen
in our likelihood fit.
Independent of the scale $M_M$ one can see that the two heaviest
neutrino masses are always very close together on a logarithmical
scale. We have to consider that a statistical accident in our scheme.

The analysis of neutrino mixing angles also is not very definitive.
Neutrino mixing is somewhat different than quark mixing. Let us
review the experimental data and the mixing formalism within the
context of the seesaw mechanism \cite{Ramond}.

The lepton equivalent of the CKM matrix is called the MNS matrix. We can
parameterize the MNS matrix in the form \cite{PDG}
\begin{equation}
V_{MNS} \! =  \! \left( \! \!
      \begin{array}
       {*{3}{c}}
       c_{12}c_{13} & s_{12}c_{13} & s_{13}e^{-i\delta} \\
       -s_{12}c_{23} - c_{12}s_{23}s_{13}e^{i\delta} & c_{12}c_{23} - s_{12}s_{23}s_{13}e^{i\delta} & s_{23}c_{13} \\
       s_{12}s_{23}-c_{12}c_{23}s_{13}e^{i\delta} & -c_{12}s_{23} - s_{12}c_{23}s_{13}e^{i\delta} & c_{23}c_{13}\\ .
      \end{array} \! \right) \! \times \hspace*{1pt} \text{diag} \hspace*{-2pt} \left(\hspace*{-1pt} e^{i\alpha_{1}/2},e^{i\alpha_{2}/2},1 \hspace*{-1pt} \right)\label{mns}
\end{equation}
where $c_{ij} \equiv \cos \theta_{ij}$ and $s_{ij} \equiv \sin \theta_{ij}$.
The three $\theta_{ij}$'s are mixing angles and $\delta, \alpha_{1}, \alpha_{2}$
are CP violating phases. The phases $\alpha_i$ are only observable
if neutrinos are Majorana particles.
The present $3 \sigma$ allowed ranges for the mixing angles are \cite{neutrino, bahcall}
\begin{equation*}
 \sin^2 2\theta_{23} \geq  0.92
\end{equation*}
\begin{equation*}
 0.70 \leq \sin^2 2\theta_{12} \leq 0.95
\end{equation*}
\begin{equation}
 \sin^2 \theta_{13} \leq  0.048.
\end{equation}
and nothing is known so far about the CP violating phases.
The existence of the two large mixing angles $\theta_{12}$ and
$\theta_{23}$ is the most striking difference between the MNS and
the CKM matrix.

Both the mass matrices for charged leptons and for neutrino have
to be diagonalized and their lefthanded unitary rotation matrices
give rise to the MNS matrix. For the charged leptons the
diagonalization procedure is analogous to the ones for the quarks:
\begin{equation}
     m^{(l)} = \left(
      \begin{array}
       {*{3}{c}}
       m_e & 0 & 0 \\
       0 & m_\mu & 0 \\
       0 & 0 & m_\tau \\
      \end{array} \right) = V_L^{(l) \dagger} M_0^{(l)} V_R^{(l)}
\end{equation}
 After the seesaw mechanism the neutrino mass matrix is
  \begin{equation} \label{seesawmatrix}
   M_0^{(Seesaw)} = M_0^{(D)} \frac {1} {M_0^{(Maj)}} M_0^{(D) T}
  \end{equation}
where all ingredients are $3 \times 3$ matrices.  Note that $M_0^{(Maj)}$
has to be symmetric. The Dirac part is diagonalized as follows:
    \begin{equation}
     m^{(D)} = \left(
      \begin{array}
       {*{3}{c}}
       m_{D,1} & 0 & 0 \\
       0 & m_{D,2} & 0 \\
       0 & 0 & m_{D,3} \\
      \end{array} \right) = V_L^{(D) \dagger} M_0^{(D)} V_R^{(D)}
    \end{equation}
    Plugging the diagonalization of the Dirac part into
    Eq. (\ref{seesawmatrix}) yields
    \begin{equation}
     M_0^{(Seesaw)}  = V_L^{(D)} \ \mathcal C \ V_L^{(D) T}
    \end{equation}
    where the central matrix $\mathcal C$ is defined as
    \begin{equation} \label{centralmatrixdef}
     \mathcal C = m^{(D)} V_R^{(D) \dagger}
     \frac {1} {M_0^{(Maj)}} V_R^{(D) *} m^{(D)}.
    \end{equation}
    The central matrix $\mathcal C$ is diagonalized with a unitary matrix $\mathcal F$:
    \begin{equation}
     \mathcal C = \mathcal F \ m_\nu \ \mathcal F^T = \mathcal F \left(
      \begin{array}
       {*{3}{c}}
       m_{1} & 0 & 0 \\
       0 & m_{2} & 0 \\
       0 & 0 & m_{3} \\
      \end{array} \right) \mathcal F^T
    \end{equation}
   The masses in the diagonal matrix $m_\nu$ are the physical neutrino masses.

The MNS matrix involves the rotations that diagonalize the mass
matrices of the charged leptons and the neutrinos. This also includes the
rotation that diagonalizes the central matrix. Therefore, in terms of the
quantities defined above, the MNS matrix becomes
\begin{equation}
     V_{MNS}=V_L^{(l) \dagger} V_L^{(D)} \mathcal F. \label{equationmns}
\end{equation}

One might hope that the striking features of the neutrino mixing
matrix might allow us to gain some insight into the Majorana sector
of the theory, which comes from very high scale physics and for
which we cannot argue for any preferred weight. Unfortunately this
does not happen. We have simulated the neutrino mixing by
considering a variety of possibilities for the Majorana mass matrix.
We have allowed for different power law weights for the Majorana
mass matrix as well as flat distributions or even a common mass. In
each case we took the Dirac Yukawa couplings to be distributed with
the weight $\sim 1/m^{1.16}$. The results for the mixing angles in
each case were so similar that we do not bother to display the
differences. In each case, the mixing angles were dominated by small
mixings, as we also found for the CKM mixing angles.

The reason for that is that the only difference within our scheme
between the CKM and the MNS matrix is the additional factor of the
rotation matrix $\mathcal F$ that contributes to the MNS matrix.
$\mathcal F$ diagonalizes the central matrix $\mathcal C$ from Eq.
(\ref{centralmatrixdef}) that contains two factors of $m^{(D)}$.
Since the diagonal Dirac mass matrices $m^{(D)}$ come from a
distribution $\rho \sim 1/m^\delta$ we usually expect the entries of
$m^{(D)}$ to have a hierarchical pattern with $m_{D,1} \ll m_{D,2}
\ll m_{D,3}$. In order to obtain large mixing angles from $\mathcal
F$ some of the entries in the central matrix $\mathcal C$ must be of
the same order. That however can only be achieved through a
numerical coincidence or a correlated hierarchy of $M_0^{(Maj)}$
designed to counterbalance the hierarchy present in $m^{(D)}$
\cite{Ramond}. Our random matrices $M_0^{(Maj)}$ are always
simulated independently from all Dirac mass terms and thus there is
no correlated hierarchy and not many more cases of two large mixing
angles than for the CKM matrix, no matter how we simulate the
Majorana mass matrix.

From now on, we always use Majorana mass matrices with elements distributed
uniformly between zero and a Majorana scale $\Lambda_{Maj}$. The charged
lepton mass matrices and the Dirac mass matrices are simulated using
a weight $\rho \sim 1/m^{1.16}$. The neutrino masses we get from these
simulations are close to a scale invariant distribution similar to the
case of the quarks as seen in Fig. \ref{eigenvalues_delta_1.16_logscale}
in Section 4.

We are interested in the probability of finding large mixing angles,
especially the ``two large mixing angle'' solution found in nature,
using the parameterization of Eq. (\ref{mns}). We will define a
large mixing angle to be an angle between $30^o $ and $60^o$ which
corresponds to $\sin^2 2\theta > 0.75$.

In studying the properties of neutrino mixing, we have run
simulations under various conditions. In the first setting, we
impose no constraints at all and compare the percentage of two large
mixing angles between the CKM matrix and the MNS matrix. We find two
large mixing angles in 5\% of the simulations for the CKM matrix,
whereas for MNS matrix the percentage is a little higher being
6.5\%. This difference is due to the seesaw mechanism i.e. due the
effect of $\mathcal F$ matrix in Eq. (\ref{equationmns}). Note that
the resulting distributions of mixing angles for the MNS matrix are
independent of the Majorana scale when there are no constraints.

We also have run simulations with constraints on the charged lepton masses
and the neutrino masses. For the charged lepton masses we adopt similar
constraints as we did in the previous section for the quarks considering
only those mass matrices for which the mass eigenvalues for the electron
($i.e$ the lightest eigenvalue of the charged lepton matrix) are less than 5
MeV, the mass eigenvalues for the muon are within 25 MeV and 1 GeV and
the mass eigenvalues for the tau are greater than 1.5 GeV.
For the neutrino masses we explore two different kinds of conditions.
In one set of constraints, we accept only those neutrino masses where the
mass differences lie within the experimental $3\sigma$ allowed ranges and
the cosmological bound is satisfied.
A second set of simulations only requires that the two heaviest masses lie in
the two decade wide mass range
$0.001 ~{\rm eV}< m_{2},~ m_{3} < 0.1~{\rm eV}$ with a mass ratio
$m_{3}/m_{2} < 10$, where $m_{3}$ and $m_{2}$ are the heaviest and
second heaviest neutrinos respectively. This is more general than the
first set of neutrino mass constraints, but captures what could be
important features of the physical masses. In both the constrained
cases, the number of cases with two large mixing angles depends on the
Majorana scale.

\begin{table}[h]
 \begin{center}
  \begin{tabular}{|c|c|c|c|c|c|} \hline
   $\Lambda_{Maj} \text{ [GeV]}$ & $10^{14}$  & $10^{12}$ & $10^{11}$  & $10^{9}$  & $10^{7}$  \\ \hline
   Zero LMA & $63.1 \%$ & $ 60.2 \%$& $ 57.6 \%$ & $53.2 \%$
   & $40.3 \%$ \\ \hline
   One LMA & $31.0 \%$ & $ 32.8 \%$& $ 34.0 \%$ & $35.9 \%$
   & $40.7 \%$ \\ \hline
   \bf{Two LMA} & $4.5 \%$ & $5.4 \%$ & $6.6 \%$ & $8.7 \%$
   & $15.4 \%$ \\ \hline
   Three LMA & $1.5 \%$ & $1.6 \%$ & $1.8 \%$ & $2.2 \%$
   & $3.7 \%$ \\ \hline
  \end{tabular}
  \caption{\small{Percentages of outcomes depending on the number of large mixing angles and the Majorana scale when the masses satisfy the 3$\sigma$ bounds for the neutrino mass differences and the cosmological bound.}}
  \label{lma change}
 \end{center}
\end{table}

\begin{table}[h]
 \begin{center}
  \begin{tabular}{|c|c|c|c|c|c|} \hline
   $\Lambda_{Maj} \text{ [GeV]}$ & $10^{14}$  & $10^{12}$  & $10^{11}$  & $10^{9}$  & $10^{7}$ \\ \hline
   Zero LMA & $49.5 \%$ & $47.8 \%$ & $ 46.4 \%$ & $42.4 \%$
   & $33.1 \%$ \\ \hline
   One LMA & $42.9 \%$ & $43.2 \%$ & $ 43.2 \%$ & $43.4 \%$
   & $44.2 \%$ \\ \hline
   \bf{Two LMA} & $5.8 \%$ & $6.8 \%$& $8.0 \%$ & $11.1 \%$
   & $17.7 \%$ \\ \hline
   Three LMA & $1.8 \%$ & $2.2 \%$ & $ 2.5 \%$ & $3.0 \%$
   & $5.0 \%$ \\ \hline
  \end{tabular}
  \caption{\small{Percentages of outcomes depending on the number of large mixing angles and the Majorana scale when the masses $m_{2}$ and $m_{3}$ are in the range $0.001-0.1$ eV with a mass ratio $m_{3}/m_{2} < 10$.}}
  \label{lma change2}
 \end{center}
\end{table}

In Tables \ref{lma change} and \ref{lma change2} we present the
results for the simulations concerning the number of large mixing
angles when the above constraints are imposed. We see that the
probability of two large mixing angles is never dominant. However,
this probability is not small either. Depending on the scenario
considered and the Majorana scale, the percentage ranges from about
4\% to 18\% as we vary $\Lambda_{Maj}$. There is a general trend in
the cases where constraints on the masses are imposed - the
percentage is larger when the Majorana scale is smaller. This is
readily understandable. When the Majorana scale is small, the
average element in the mass matrix is comparable to or greater than
the resulting mass eigenvalues, because in the seesaw mechanism the
masses are proportional to $1/\Lambda_{Maj}$. When all elements are
large, one requires larger mixing angles to diagonalize the mass
matrix.

Our results are somewhat inconclusive. One finds solutions with two
large mixing angles with a reasonable percentage. However, it is not
the most likely outcome. Statistically, we can not conclude anything
significant from a comparison with the observed mixing angles. This
conclusion is independent of how we have modeled the possible
randomness of the Majorana sector. We have also found that the
information on the neutrino mass differences is unable to tightly
constrain the scale of the Majorana sector.

We can also extract some predictive elements from our analysis. We
will study these using a Majorana scale of $\Lambda_{Maj} = 10^{12}
\text{ GeV}$. In each case we also only consider those realistic
solutions that satisfy the $3 \sigma$ bounds on the neutrino mass
differences as well as the cosmological bound and that have large
angles $\theta_{12}$ and $\theta_{23}$ and values of $\theta_{13}$
smaller than the experimental bound, and thus are very similar to
the real data.

\begin{figure}[ht]
 \begin{center}
  \includegraphics[scale=0.75]{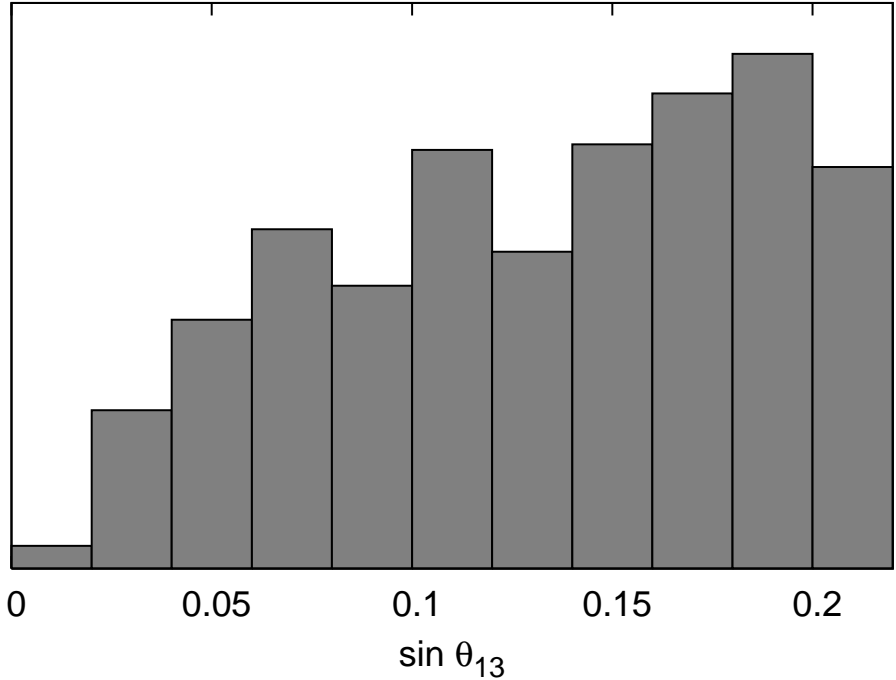}
 \end{center}
 \caption{\small{The distribution of $\sin \theta_{13}$}}
 \label{theta13_constrain}
\end{figure}

Using these realistic simulations, we have studied the size of the
third mixing angle $\theta_{13}$. The distribution of $\sin
\theta_{13}$ is shown in Fig. \ref{theta13_constrain} within its $3
\sigma$ allowed range, that is for $\sin^2 \theta_{13} \leq 0.048$.
We see that the distribution of $\sin \theta_{13}$ is not peaked at
low values but rises towards its $3\sigma$ upper bound, so that
there is no reason to expect that the third mixing angle is
suppressed. The median of the distribution is $0.14$ and $95 \%$ of
the values lie above 0.04.

\begin{figure}[ht]
 \begin{center}
  \includegraphics[scale=0.75]{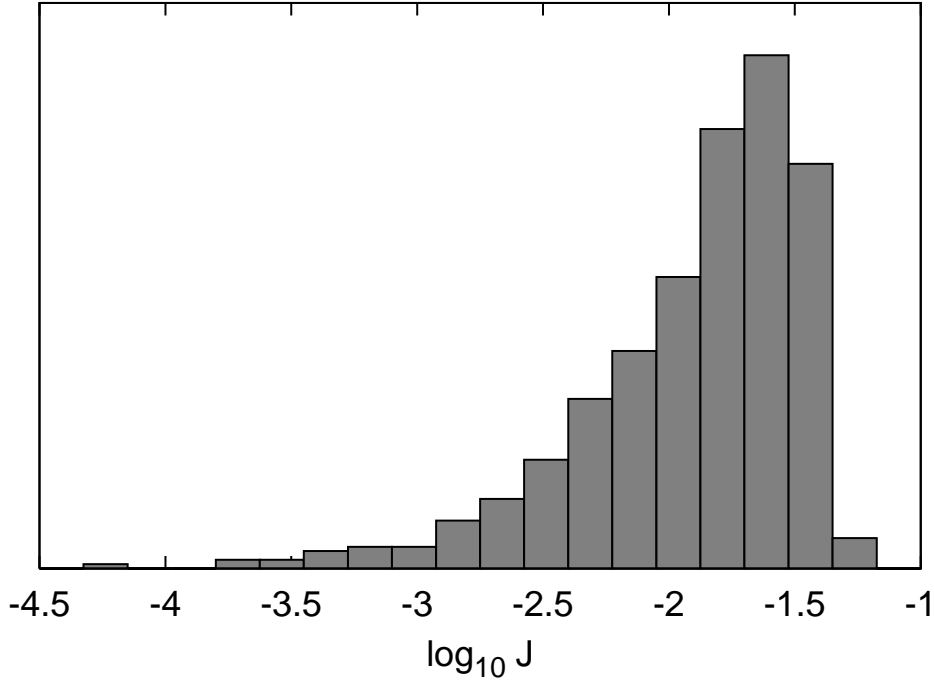}
 \end{center}
 \caption{\small{Distribution of the Jarlskog invariant $J$ for neutrinos}}
 \label{J_Neutrinos}
\end{figure}

We also have studied the neutrino equivalent of the Jarlskog
invariant that signals the strength of CP violation for lepton
number conserving processes the neutrino sector. This is plotted in
Fig. \ref{J_Neutrinos}. Here we see that the Jarlskog invariant is
generally large, of order $10^{-2}$. We see that the peak of the
distribution is close to its theoretical maximum possible value
which is $1/(6\sqrt{3})$. The median of the distribution is at 0.016
and 95\% of the values are above $1.6\times 10^{-3}$.

\begin{figure}[ht]
 \begin{center}
  \includegraphics[scale=0.75]{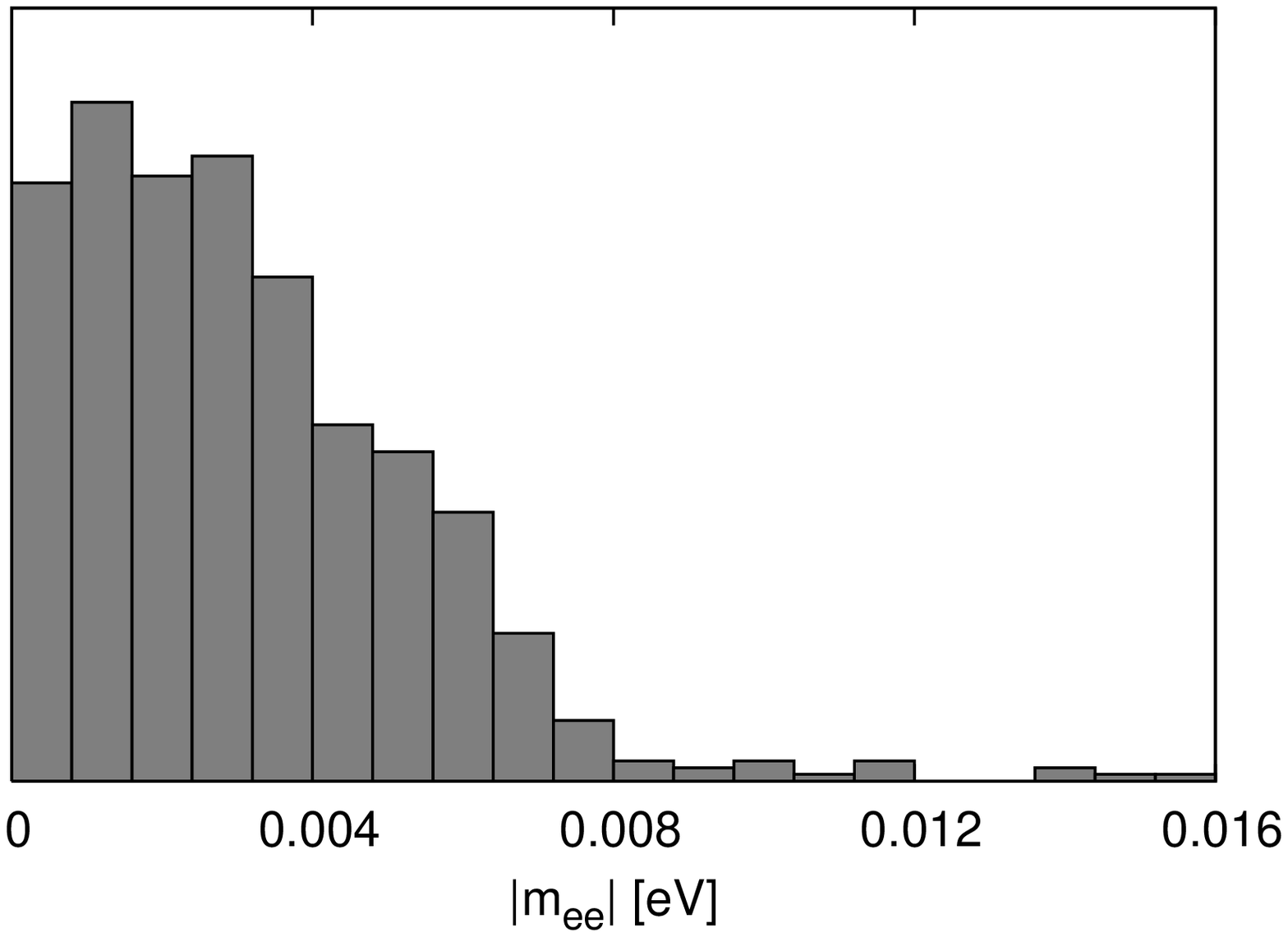}
 \end{center}
 \caption{\small{Distribution of the effective Majorana mass $m_{ee}$}}
 \label{mee}
\end{figure}

Finally we consider the effective Majorana mass defined by
\begin{equation}
 m_{ee} = \sum_i U_{ei}^2 ~m_i
\end{equation}
which is the figure of merit in neutrinoless double beta decay
\cite{bilenky}. The simulations of this parameter are shown in Fig.
\ref{mee} again for the realistic case very similar to nature. We
see that the distribution decreases approximately linearly until a
value of 0.008 eV, with only very few values higher than that. These
come from cases where the neutrino masses are not strictly
hierarchical and the lightest mass is close to the second lightest
mass. The median of the distribution is at 0.0028 eV, 95\% are
higher than $3.5 \times 10^{-4}$ eV. 95\% of the values lie below
0.0068 eV which is consistent with other results that expect the
effective Majorana mass for the case of hierarchical neutrino masses
to be smaller than 0.0064 eV \cite{bilenky}.

A pioneering effort in simulating neutrino mixing angles randomly in
the context of the seesaw mechanism was the work of Goldman and
Stephenson in 1981 \cite{goldman}. These authors randomly populated
both the Dirac and Majorana mass matrices with a uniform
distribution and found that the typical case involved angles of the
same size as the Cabibbo angle. The intent was not to invoke a
random dynamical framework but to argue that unless the initial
Yukawa couplings were peculiar, that we should expect to find
Cabibbo-like mixing angles. More recently, the idea has surfaced
under the label of ``anarchy'' in the works of \cite{anarchy}. These
authors build in more of what is now known about the neutrino mass
spectrum and, although again one most typically finds Cabibbo-like
angles, they point out that one finds two large mixing angles
(defined as $\sin^2 2\theta >0.5$) in about 13\% of the cases when
simulated with a uniform random distribution. Our considerations
have been somewhat different. We have used the Dirac component of
the neutrino masses to be the same distribution that has been used
in the description of the charged leptons and the quarks, and
considered various possibilities for the Majorana sector. However,
our numerical results are fairly similar. Two large mixing angles
are found between 5\% and 15\% of the time as we change
$\Lambda_{Maj}$ and when we consider output masses similar to those
needed experimentally. Note that the definition of ``large mixing
angles'' is more generous in the anarchy papers ($\sin^2 2\theta
>0.5$) than in our work ($\sin^2 2\theta >0.75$). In our simulations
this can make a factor of two difference in the percentage of cases
that are said to have two large mixing angles. For example, with
$\Lambda_{Maj}= 10^{12}$~GeV, the percentage of two large mixing
angle solutions goes from 5.4\% to 10.1\% when adopting the more
generous range. However, this does not change our conclusions. Large
mixing angles are found a significant fraction of the time, but are
not the most probable solution.

Our work has concentrated on the standard seesaw mechanism. If
considering the Type II seesaw~\cite{type2} or triplet
seesaw~\cite{triplet} mechanisms instead, the neutrino mass matrix
is completely determined by physics at a very high scale for which
we have no basis to prefer a certain weight. This leads to more
freedom because of the lack of constraints, but unfortunately also
to no predictive power.

\section{Summary and discussion}

We have provided an exploration of the distribution of quark and
lepton masses and mixings, treating these as random variables
distributed with respect to some probability distribution function,
which we called the weight. We have shown that this weight is close
to being scale invariant in form, i.e $\rho(m)\sim1/m$. In
particular a statistical analysis of power law weights
$\rho\sim1/m^\delta$ tightly constrains the power to be $\delta =
1.02\pm 0.08$. Such weights commonly produce small mixing angles and
a hierarchy of CKM elements. Our exploration of the neutrino sector
was somewhat inconclusive. The use of the usual seesaw mechanism,
combined with the expected weight for the Dirac masses, could
marginally accommodate the observed mass differences and mixing
angles, but the existence of large mixing angles was not a common
occurrence in this framework. However, the neutrino sector does
allow us to make statistical predictions of this framework. A normal
hierarchy of neutrino masses is highly favored by the scale
invariant weights. We saw that the leptonic Jarlskog invariant is
typically of order $10^{-2}$ and with 95\% confidence greater than
$1.6 \times 10^{-3}$. The third presently unmeasured MNS mixing
angle $\sin \theta_{13}$ is typically of order 0.1 and with 95\%
confidence it is larger than 0.04. The effective Majorana mass
$m_{ee}$ is typically of order 0.001 eV.

We are aware of many caveats about this procedure, and there are
probably others that we do not enumerate. Some of these issues are:

1) Differences between quarks and leptons: The information that we
have about the masses exists at low energy. However, it is likely
that the fundamental input consists of parameters defined at a high
energy scale. We explored the effect of this using only the Standard
Model to run the Yukawa couplings to the GUT scale, and this did not
modify any of our conclusions. However, we clearly do not know the
physics at intermediate scales. If the intermediate physics only
produces logarithmic running, the residual effect is also probably
not large.

2) Generation structure: The assumption of statistical independence
of all masses may fail in some obvious ways. The most suspect aspect
is that we do not assume any correlation between members of the same
generation. In some grand unified theories, this is incorrect,
leading for example to a relation between the $\tau$ and $b$ quark
masses \cite{tau}. Perhaps in the fundamental theory the random
distribution occurs separately for the average value of a
generation's masses and for the ratio of masses. This pattern is
distinct from that which we have assumed.

3) Anthropic constraints: In a landscape picture there are
inevitable anthropic constraints. Some parameter sets that occur in
the landscape do not allow the existence of life, and hence we could
not find ourselves in a portion of the multiverse where these
parameters occurred. These constraints may distort the mass
distribution. We have provided an estimate of this effect in the
discussion of quark and lepton masses, and it appears not to change
our qualitative conclusions. We have not considered a possible
anthropic constraint on neutrino masses~\cite{teg}, which would
disfavor masses above 1~eV. In general, it is hard to fully explore
the effect of anthropic constraints without having the full
fundamental theory.

4) Other flavor symmetries: Many of the attempts to understand quark
and lepton masses have focussed on adding extra flavor symmetries to
the Standard Model. Our basic assumption asserts that vacua with
these extra symmetries do not constitute a large part of the string
theory landscape of vacua.

From these comments, it is clear that our investigation is a rather
preliminary attempt at phenomenology in the context of the
landscape. In the end, the only fully compelling procedure is to
identify the fundamental theory, such as string theory, that leads
to the landscape and fully solve it. With this solution in hand one
can investigate directly those vacua that lead to the Standard Model
and calculate the statistical features of those solutions. Then one
can directly address the constraints and symmetries that occur due
to intermediate scale physics and can assess whether the fundamental
theory is statistically compatible with the data. However, such a
``top down'' solution is clearly very difficult and we are far from
being able to accomplish it. In the meantime we need to tentatively
explore the phenomenology from the ``bottom up'' as best we are
able.

It is hoped that eventually this weight can be calculated from a
more fundamental theory such as string theory. An illustrative
example was given in Ref. \cite{Mdynamics}, using the Intersecting
Brane Worlds (IBW) construction of the Standard Model \cite{ibw}.
There, the Yukawa couplings arise from the a non-perturbative effect
spanning the area between the intersections of three pairs of branes
($A_{ijk}$), with the Yukawa interactions being exponentially
suppressed in that area,
\begin{equation}
\Gamma_{ijk} = \Gamma_0 e^{-\frac{A_{ijk}}{2\pi\alpha'}}
\end{equation}
up to a phase. In this case a flat distribution in the area
$\rho(A)\sim\rm{constant}$ yields the scale invariant weight in
masses. The range in the allowed values of the area translates into
a finite range in the masses. It is possible that this exponential
behavior could be a more general feature of the landscape. Of
course, this relation does not fully explain the nearly scale
invariant distribution in the masses, but only transfers the problem
to the string scale to understand whether the distribution in the
area is nearly flat. It would be interesting to explore possible
dynamics that could produce an ensemble of Standard Model string
vacua and assess the resulting effect on the distribution of masses.

\section*{Acknowledgement}
We would like to thank Guy Blaylock for valuable discussions.
This work has been supported in part by the U.S. National Science
Foundation.

\section*{Appendix:  The weight for \(\Delta m^2\) and the Majorana
scale}
 When we consider the neutrino masses by themselves without mixing,
 the light neutrino masses $m_i$ are of the form
 \begin{equation}
  m_i = \frac{m_{D,i}^2}{M_M}
 \end{equation}
 where we take $M_M$ as a common scale. Using the scale invariant weight
 \begin{equation} \label{rhodeltaonemD}
  \rho \left( m_{D,i} \right) = \frac {1} {\log \frac {m_*} {m_{low}}} \ \frac {1} {m_{D,i}} \ \Theta
   \left( m_{D,i} - m_{low} \right) \Theta \left( m_* - m_{D,i} \right)
 \end{equation}
 for the Dirac masses, we derive the weight for the
 neutrino mass differences \(\Delta m^2_{ij} = m_j^2 - m_i^2\) and use
 the physical data for a likelihood fit of the Majorana scale $M_M$.

 As a first step let we calculate the weight for the neutrino masses \(m_i\) with the scale
 invariant random Dirac masses \(m_{D,i}\):
 \begin{align} \label{pdfnetrinomasses}
  \rho \left(m_i\right) & = \int dm_{D,i} \ \rho \left(m_{D,i}\right) \ \delta
                             \left(\frac {m_{D,i}^2} {M_M} - m_i\right) \notag \\
                        & = \frac {1} {2 \log \frac {m_*} {m_{low}}} \ \frac {1} {m_i} \
                             \Theta \left(m_i - \frac {m_{low}^2} {M_M}\right)
                             \Theta \left(\frac {m_*^2} {M_M} - m_i\right) \notag \\
                        & = \frac {1} {\log \frac {M_U} {M_L}} \ \frac {1} {m_i} \
                             \Theta \left(m_i - M_L\right)
                             \Theta \left(M_U - m_i\right)
 \end{align}
 where we defined \(M_L = \frac {m_{low}^2} {M_M}\) and
 \(M_U = \frac {m_*^2} {M_M}\) for simplicity.
 We see that the neutrino masses are also distributed with respect to a scale invariant weight.
 The lower and upper bounds are of course changed as we would expect and
 the result in Eq. (\ref{pdfnetrinomasses}) is automatically normalized since we started from a
 normalized weight.

 As a next step we calculate the weight for the neutrino masses squared,
 \(\rho \left(m_i^2\right)\). The calculation is analogous to the one for
 \(\rho \left(m_i\right)\) in Eq. (\ref{pdfnetrinomasses}) and the result is
 \begin{align}
  \rho \left(m_i^2\right) & = \frac {1} {\log \frac {M^2_U} {M^2_L}} \ \frac {1} {m_i^2} \
                             \Theta \left(m_i^2 - M^2_L\right)
                             \Theta \left(M^2_U - m_i^2\right).
 \end{align}

 With that we will calculate the weight for \(\Delta m^2_{ij} = m_j^2 - m_i^2\):
 \begin{align}
  \rho \left(\Delta m^2_{ij}\right) & = \int dm_i^2 \int dm_j^2 \ \rho \left(m_i^2\right)
                                          \rho \left(m_j^2\right) \delta \left(m_j^2 - m_i^2
                                          - \Delta m^2_{ij}\right) \notag \\
                                      & = \frac {1} {\left(\log \frac {M^2_U} {M^2_L}\right)^2}
                                          \, \frac {1} {| \Delta m^2_{ij} |} \
                                          \log \frac {\left(M^2_U - |\Delta m^2_{ij}| \right)
                                                      \left(M^2_L + |\Delta m^2_{ij}|\right)}
                                                     {M^2_U \, M^2_L}
 \end{align}
 As we would expect,
 \(\rho \left(\Delta m^2_{ij}\right) = \rho \left(-\Delta m^2_{ij}\right)
   = \rho \left(|\Delta m^2_{ij}|\right)\), meaning that there is no preferred sign for
 \(\Delta m^2_{ij}\). The weight for \(\Delta m^2_{ij}\) is basically proportional to one over
 \(\Delta m^2_{ij}\) up to some logarithmic corrections. For the complete weight
 \(\rho \left(\Delta m^2_{ij}\right)\), we need to include the range of possible values of
 \(\Delta m^2_{ij}\) with a Theta function:
 \begin{equation} \label{pdfdeltamsquared}
  \rho \left(\Delta m^2_{ij}\right) \hspace*{-1pt} = \hspace*{-1pt} \frac {1} {\left(\log \frac {M^2_U} {M^2_L}\right)^2} \,
                                        \frac {1} {| \Delta m^2_{ij} |} \, \log \! \frac
                                        {\left(M^2_U \! - \! |\Delta m^2_{ij}| \right) \! \hspace*{-0.75pt}
                                        \left(M^2_L \! + \! |\Delta m^2_{ij}|\right)}
                                        {M^2_U \, M^2_L} \, \Theta \!
                                        \left(M^2_U \! -\! M^2_L \! - \! |\Delta m^2_{ij}|\right)
 \end{equation}
 Again, we find that our result is automatically normalized.
 Recall that we defined \(M_L = \frac {m_{low}^2} {M_M}\) and
 \(M_U = \frac {m_*^2} {M_M}\) so that \(\rho \left(\Delta m^2_{ij}\right)\)
 depends on the Majorana scale \(M_M\) as well as on \(m_{low}\) and \(m_*\).
 Our analytic result for $\rho \left(\Delta m^2_{ij}\right)$ agrees exactly
 with the result of corresponding simulations.

 One benefit of explicitly knowing the weight of an observable is that one can perform a
 likelihood fit with it. If we used the weight for \(\Delta m^2\) along with the
 physical data for $\Delta m^2_{12}$ and $\Delta m^2_{13}$ to fit the
 Majorana scale $M_M$ using a likelihood function
 $L = \rho \left(\Delta m^2_{12}\right) \rho \left(\Delta m^2_{13}\right)$,
 we would make a mistake by assuming the two measured mass differences are independent
 of each other. Therefore we have to first calculate the weight for two neutrino mass
 differences $\rho \left(\Delta m^2_{ij}, \Delta m^2_{ik}\right)$ that takes
 the correlation into account:
 \begin{align}
  \rho \left(\Delta m^2_{ij}, \Delta m^2_{ik}\right) & = \int dm_i^2 \int dm_j^2 \int dm_k^2 \ \rho \left(m_i^2\right)
                                          \rho \left(m_j^2\right) \rho \left(m_k^2\right) \notag \\
                                      & {}\hspace*{121pt}  \delta \left(m_j^2 - m_i^2 - \Delta m^2_{ij}\right)
                                          \delta \left(m_k^2 - m_i^2 - \Delta m^2_{ik}\right) \notag \\
                                      & = \frac {1} {\left(\log \frac {M^2_U} {M^2_L}\right)^3}
                                          \int dm_i^2 \ \frac {1} {m_i^2} \,
                                          \frac {1} {m_i^2 + \Delta m^2_{ij}} \,
                                          \frac {1} {m_i^2 + \Delta m^2_{ik}} \notag \\
                                      &  {}\hspace*{105pt} \Theta \left(m_i^2 - M^2_L\right)
                                          \Theta \left(M^2_U - m_i^2\right) \notag \\
                                      &  {}\hspace*{105pt} \Theta \left(m_i^2 - (M^2_L\ - \Delta m^2_{ij})\right)
                                          \Theta \left((M^2_U - \Delta m^2_{ij}) - m_i^2\right) \notag \\
                                      &  {}\hspace*{105pt} \Theta \left(m_i^2 - (M^2_L\ - \Delta m^2_{ik})\right)
                                          \Theta \left((M^2_U - \Delta m^2_{ik}) - m_i^2\right) \notag \\
                                      & = \frac {1} {\left(\log \frac {M^2_U} {M^2_L}\right)^3}
                                          \Bigg[\frac {1} {\Delta m^2_{ij} \, \Delta m^2_{ik}} \,
                                          \log \frac{M_U^2 - \Delta m^2_{ik}} {M_L^2} \notag \\
                                      & {} \hspace*{68pt} + \frac {1} {\Delta m^2_{ij} \left(\Delta m^2_{ij} - \Delta m^2_{ik}\right)} \,
                                          \log \frac{M_U^2 - \Delta m^2_{ik} + \Delta m^2_{ij}} {M_L^2 + \Delta m^2_{ij}} \notag \\
                                      & {} \hspace*{68pt} + \frac {1} {\Delta m^2_{ik} \left(\Delta m^2_{ik} - \Delta m^2_{ij}\right)} \,
                                          \log \frac{M_U^2} {M_L^2 + \Delta m^2_{ik}} \Bigg] \notag \\
                                      & {} \hspace*{68pt}
                                        \Theta \! \left(\Delta m^2_{ik}\right) \Theta \!\left((M^2_U \! -\! M^2_L) \! - \! \Delta m^2_{ik}\right) \notag \\
                                      & {} \hspace*{68pt}
                                        \Theta \! \left(\Delta m^2_{ik}\right) \Theta \! \left(\Delta m^2_{ik} - \! \Delta m^2_{ij}\right) + \dots
 \end{align}
 Here we only display the part of the result for the case where
 $0 < \Delta m^2_{ij} < \Delta m^2_{ik}$ which is the case for a normal neutrino mass
 hierarchy when we identify $\Delta m^2_{ij}$ with $\Delta m^2_{12}$ and
 $\Delta m^2_{ik}$ with $\Delta m^2_{13}$. Again, this weight depends
 on the Majorana scale $M_M$ via $M_L$ and $M_U$.

 Now we can finally perform a likelihood fit for the Majorana scale $M_M$
 using the likelihood function
 $L\left(M_M\right) = \rho \left(\Delta m^2_{12}, \Delta m^2_{13}\right)$.
 As the experimental input we use the cental values \cite{LP2005}
 $\Delta m^2_{12} = 8.0 \times 10^{-5} \text{ eV}^2$ and
 $\Delta m^2_{13} = 2.3 \times 10^{-3} \text{ eV}^2$.
 We find that $M_M$ has to be smaller than $10^{15} \text{ GeV}$
 to accommodate the experimental data. As a preferred value the fit
 yields $9 \times 10^{14} \text{ GeV}$. At the 1\(\sigma\) level
 \(M_M < 3 \times 10^{10} \text{ GeV}\) are excluded and at the 2\(\sigma\)
 level \(M_M < 10^{5} \text{ GeV}\) are excluded.

 To interpret these results it is useful to have a second look at
 Fig. \ref{neutrinos} where we assume hierarchical neutrino masses
 $m_1 \ll m_2 \ll m_3$. For high $M_M$ there is more ``phase space'' for
 the lightest neutrino mass which explains the preferred value close to its
 upper limit. But since this ``phase space'' decreases only linearly
 when one goes to smaller $M_M$ on a logarithmic scale the the range of
 allowed Majorana scales $M_M$ spans many orders of magnitude.

\end{document}